\documentclass[
 aip,
 cha,
 amsmath,amssymb,
 reprint,
 twocolumn,
 floatfix,
]{revtex4-2}

\usepackage[dvipsnames]{xcolor}
\usepackage{graphicx}
\usepackage{dcolumn}
\usepackage{bm}
\usepackage{xr-hyper}
\usepackage[hidelinks]{hyperref}

\newcommand{\ee}[1]{\mathrm{e}^{#1}}
\renewcommand{\vec}[1]{\mathbf{#1}}
\newcommand{\CRPS}[1]{\mathrm{CRPS}^{#1}}
\newcommand{\RMSE}[1]{\mathrm{RMSE}^{#1}}
\newcommand{\SPRD}[1]{\mathrm{SPRD}^{#1}}

\newcommand{\ens}[1]{\mathbf{x}^{#1}}
\newcommand{\ensmean}[1]{\overline{\mathbf{x}}^{#1}}
\newcommand{\ensspan}[1]{\mathbf{X}^{#1}}

\newcommand{\obs}[1]{\mathbf{y}^{#1}}
\newcommand{\obsmean}[1]{\overline{\mathbf{y}}^{#1}}
\newcommand{\obsspan}[1]{\mathbf{Y}^{#1}}

\newcommand{\unc}{\eta}

\makeatletter
\newcommand*{\addFileDependency}[1]{
  \typeout{(#1)}
  \@addtofilelist{#1}
  \IfFileExists{#1}{}{\typeout{No file #1.}}
}
\makeatother
\newcommand*{\myexternaldocument}[1]{%
    \externaldocument{#1}%
    \addFileDependency{#1.tex}%
    \addFileDependency{#1.aux}%
}
\myexternaldocument{supplement}

\begin{document}

\preprint{}

\title{Reconstructing Cardiac Electrical Excitations from Optical Mapping Recordings}

\author{C. D. Marcotte}~\thanks{Corresponding Author: christopher.marcotte@durham.ac.uk}
\affiliation{Department of Computer Science, Durham University, Durham, UK, DH1 3LE}

\author{M. J. Hoffman}
\affiliation{School of Mathematical Sciences, Rochester Institute of Technology, Rochester, NY, USA, 14623}

\author{F. H. Fenton}
\affiliation{School of Physics, Georgia Institute of Technology, Atlanta, GA, USA, 30332 }

\author{E. M. Cherry}
\affiliation{School of Computational Science and Engineering, Georgia Institute of Technology, Atlanta, GA, USA, 30332 }

\date{\today}

\begin{abstract}
The reconstruction of electrical excitation patterns through the unobserved depth of the tissue is essential to realizing the potential of computational models in cardiac medicine.
We have utilized experimental optical-mapping recordings of cardiac electrical excitation on the epicardial and endocardial surfaces of a canine ventricle as observations directing a local ensemble transform Kalman Filter (LETKF) data assimilation scheme.
We demonstrate that the inclusion of explicit information about the stimulation protocol can marginally improve the confidence of the ensemble reconstruction and the reliability of the assimilation over time.
Likewise, we consider the efficacy of stochastic modeling additions to the assimilation scheme in the context of experimentally derived observation sets.
Approximation error is addressed at both the observation and modeling stages, through the uncertainty of observations and the specification of the model used in the assimilation ensemble.
We find that perturbative modifications to the observations have marginal to deleterious effects on the accuracy and robustness of the state reconstruction.
Further, we find that incorporating additional information from the observations into the model itself (in the case of stimulus and stochastic currents) has a marginal improvement on the reconstruction accuracy over a fully autonomous model, while complicating the model itself and thus introducing potential for new types of model error. 
That the inclusion of explicit modeling information has negligible to negative effects on the reconstruction implies the need for new avenues for optimization of data assimilation schemes applied to cardiac electrical excitation.
\end{abstract}


\maketitle

\begin{quotation}
Typical cardiac electrophysiology experiments record excitation data with optical cameras that are tuned to measure fluorescence of voltage-sensitive dyes and are only able to observe excitation patterns on the surfaces of tissues. 
We know from theoretical and computational evidence that the details of excitation patterns through the depth of tissues are important for prediction and control, while those observable on the surface are a very small portion of the overall excitation structure.
These patterns may be pro-arrhythmic due to structures that are unobservable near the surfaces of the tissue, which can pin organizing features of the dynamics and thus simplify the excitation patterns.
As the interiors of the tissue are inaccessible using traditional experimental means, we turn to reconstructing the interior using computational methods---specifically, data assimilation.
\end{quotation}

\section{Introduction}

Experiments in cardiac excitation dynamics provide only a limited window into a very complex system.
Micro-electrode recordings trade spatial resolution for temporal resolution, electrocardiograms eschew cellular detail for organ-level holism, and optical-mapping experiments focus on the electrical excitations near the exposed surfaces of a tissue whose internal dynamics may be much more complex.
Data assimilation (DA) promises to address the difficulties of the latter; with a reliable dynamical model and sufficiently precise observations of the experimental state, the techniques of data assimilation aim to reconstruct the unobserved interior dynamics of the tissue, subject to the observations at the surface(s).
Importantly, the techniques of data assimilation are robust with respect to uncertainty, as the uncertainty of the truth state and of the observation values, as well as of the specificity of model predictions, are all ever-present when working with real experimental data.

Data assimilation has co-developed with the weather and climate forecasting communities, for whom the tools are often specialized and well-tuned.
Ensemble methods, in particular, are relative newcomers to the application~\cite{evensen1994sequential}, but are favored due to their simpler relationship to nonlinear dynamical models.
Numerous studies have developed new data assimilation methods and compared them to existing approaches~\cite{nergerUnificationEnsembleSquare2012, Nerger2012a, todterAssessmentNonlinearEnsemble2016, Kirchgessner2017, Vetra-Carvalho2018, nergerDataAssimilationNonlinear2022}, developed innovations for improving the performance~\cite{cosmeImplementationReducedRank2010, dashPerformanceComparisonsThree2023} and robustness of results from data assimilation~\cite{raoRobustDataAssimilation2017}, defined fair scoring rules for the rigorous definition of `improving' data assimilation results~\cite{hersbachDecompositionContinuousRanked2000, brockerEvaluatingRawEnsembles2012, thoreyOnlineLearningContinuous2017, leutbecherUnderstandingChangesContinuous2021}, and even considered such subtleties as the connection between the observations, assimilation filter, and the model~\cite{leaAssessingNewCoupled2015, goodliff2019temperature, pennyStronglyCoupledData2019, tangStronglyCoupledData2021}.

Significant efforts in the assimilation of cardiac excitation patterns in three-dimensional tissue have focused on the foundational aspects of the problem---data distributions, uncertainty quantification, model error, and robustness with respect to noise~\cite{lavigne2017effects, hoffman2016reconstructing, hoffman2020sensitivity}---in the context of synthetic observations generated by a model, rather than from experiment.
In previous work~\cite{marcotte2021robust} we have found a sensitivity to some parameter uncertainties in the accuracy of local ensemble transform Kalman filter (LETKF) reconstructions, as well as improved robustness and accuracy with stochastic formulations of the underlying model.
This work considers the efficacy of our reconstruction schemes in the case of experimental data typical of the field:  periodic pacing of a tissue, with excitations captured by optical mapping on the epicardial and endocardial surfaces of the tissue.

Throughout this manuscript we will be using {a dataset } previously published in experimental studies~\cite{gizzi2013effects, loppini2019spatiotemporal, barone2020efficient, barone2020experimental}.
{The dataset corresponds } to activation patterns subject to a localized, periodic, current stimulation in canine ventricular tissue. 
We will investigate the use of both autonomous and non-autonomous models for the reconstruction of these experimental observations, and the effect of stochastic model interventions on the accuracy of the reconstructions.
Finally, we investigate the effect of additional synthetic observations generated in the interior of the tissue on the robustness of the assimilation and accuracy of the reconstruction of experimental dynamics.

\section{Methods}

In this section we outline the mathematical details of the dynamical model, the determination of reasonable model parameters, the generation of observations from the experimental data, and finally the local ensemble transform Kalman filter (LETKF) used in the assimilation of the observations for the reconstruction of the experimental dynamics.

\subsection{Model}
We use the Fenton-Karma three-variable model~\cite{fenton1998vortex} for the electrophysiology, with anisotropic diffusion controlled by a fiber orientation that rotates with in the $x,y$-plane with depth, denoted by angle with respect to the $x$-axis, $\theta(z)$.
The evolution equations are  
\begin{align}\label{eq:fk}
    \partial_t u &- \nabla\cdot (\vec{D}(\vec{x})\nabla u) = I_\mathrm{stim}-I_{fi}-I_{so}-I_{si}, \\
    \partial_t v &= \Theta(u_c-u)(1-v)/\tau_v^{-}(u) - \Theta(u-u_c)v/\tau_v^+  \\
    \partial_t w &= \Theta(u_c-u)(1-w)/\tau_w^{-} - \Theta(u-u_c)w/\tau_w^+,
\end{align}
in conjunction with the no-flux boundary conditions, $\hat{\vec{n}}\cdot\vec{D}(\vec{x})\nabla u = 0$.
The currents and explicit relaxation scale for $v$ are given by
\begin{align}\label{eq:currents}
    I_{fi}(u,v) &= -v\Theta(u-u_c)(1-u)(u-u_c)/\tau_d, \\
    I_{so}(u) &= u\Theta(u_c-u)/\tau_o + \Theta(u-u_c)/\tau_r, \\
    I_{si}(u,w) &= -w(1+\tanh(k(u-u_c^{si})))/(2\tau_{si}), \\
    \tau_v^{-}(u) &= \tau_{v1}^{-}\Theta(u-u_v) + \tau_{v2}^{-}\Theta(u_v-u),
\end{align}
save for the explicit stimulation current $I_\mathrm{stim}(t,\vec{x})$, which is zero identically for autonomous dynamics, and is specified in particular assimilation experiments (see Section~\ref{ssec:stim}).
The diffusive spatial coupling is implemented using a finite-difference scheme that follows that in the original model description~\cite{fenton1998vortex}.
The diffusive coupling of the tissue is subject to fiber orientation that rotates through the tissue depth $d$, $\theta(z) = \theta_0 + \delta\theta(-1/2 + z / d)$, where $0 \leq z \leq d$, $\theta_0$ and $\delta\theta$ are specified throughout this manuscript, with $D_{\parallel} = 0.001~\mathrm{cm}^2/\mathrm{ms}$ and $D_{\perp} = 0.0002~\mathrm{cm}^2/\mathrm{ms}$, for an (diffusion) anisotropy ratio $D_{\parallel}/D_{\perp} = 5$, which is relatively small compared to some ratios measured in cardiac tissue~\cite{spachDiscontinuousNaturePropagation1981, kotadiaAnisotropicCardiacConduction2020}.

We use the model parameters defined in Ref.~\onlinecite{barone2020experimental} for this model and experimental dataset.
These parameters are quite different from other parameter sets used with this model for physically relevant dynamics; cf. Ref.~\onlinecite{fenton1998vortex}.
However, for the present investigation we have opted to consider the smallest model errors to better determine the affect of changing the model and observation properties.

In all experiments, we use a Rush-Larson time-stepping scheme for the gating variables $v$ and $w$ combined with a Forward-Euler method for the transmembrane potential $u$~\cite{perego2009efficient}.
When we incorporate stochastic effects (cf. Section~\ref{ssec:stoch}) we adapt the Forward-Euler method for the transmembrane potential $u$ to an Euler-Maruyama method.
In this work we do not incorporate stochastic terms in the dynamics of the gating variables, as previous work demonstrated that similar ensemble inflation effectiveness can be achieved with only stochastic currents for this model~\cite{marcotte2021robust}, which makes the implementation simpler and more performant.

\subsection{Observation extraction}

The experimental data comprises optical recordings of fluorescence (c.f., Ref.~\onlinecite{gizzi2013effects, cherry2008visualization} for details of the original preparation) corresponding to transmembrane potential, which have been smoothed and denoised.
The observations ($\obs{o}$) are a vector of tuples ($\obs{o}_i$) containing a tag designating the observed field variable, coordinates in the domain ($[x_i,y_i,z_i]$), the observed field variable at these coordinates (e.g., $u_i = u(x_i,y_i,z_i)$), and the uncertainty of the observation value ($\unc_i$), for each observation.
The observed field values are also normalized such that $0 \leq u_i \leq 1$.
As the observations only cover the transmembrane potential $u$, we may use $\obs{o}_i$ and $u_i$ interchangeably to denote the values of the observations.
The recordings cover response excitations for canine ventricle tissue subject to a periodic stimulus applied from a single position, across a large range of stimulus periods or basic cycle lengths (BCLs).

\begin{figure}[htpb]
    \centering
    \includegraphics{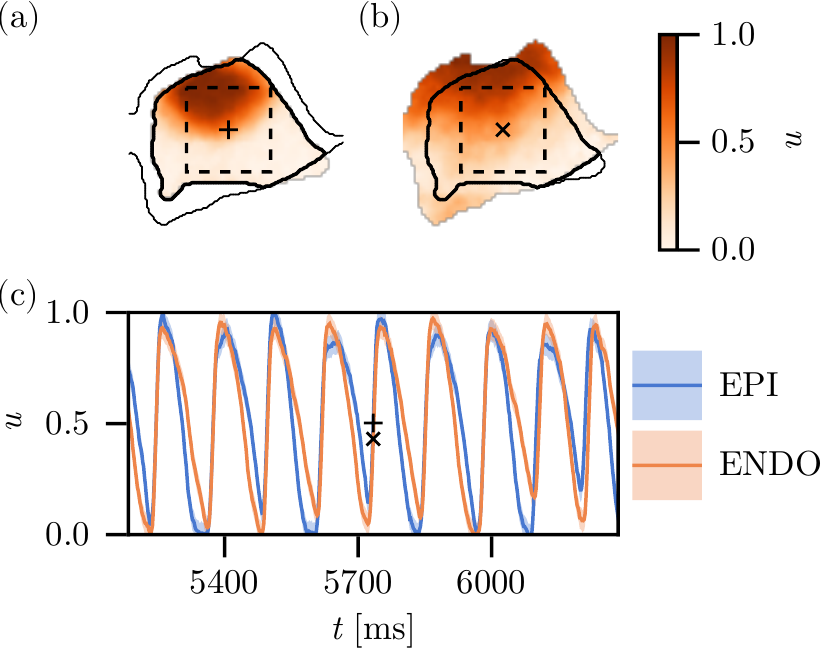}
    \caption{Example of the experimental data used for the reconstruction experiments. (a) Epicardial snapshot and (b) endocardial snapshot at time $t=5734$~ms, with (c) time traces of the recording at positions labeled by the $+$ (epicardial) and $\times$ (endocardial) markers. The thin black curve in (a) and (b) designates the boundary of the opposing surface recording, and the thick black curve designates the boundary of their mutual intersection. The inset dashed square has side-length $3$~cm, aligned to the horizontal and vertical axes, and shows the region selected to extract observations for assimilation.}
    \label{fig:datafig}
\end{figure}

For the purposes of this study, we have focused on a dataset with stimulation period of $118\pm 2$~ms{, as the patterns generated with this short stimulation period are the most irregular, spatially and temporally, and thus present the most challenging reconstruction task}.
An example of the recordings from this dataset is shown in figure~\ref{fig:datafig}, detailing (a) the epicardial and (b) the endocardial surface recordings, along with (c) temporal traces of the recording at the center-points of the tissue surfaces.
The figure likewise shows an inset $3$~cm square used for the generation of the observation sets for the assimilation experiments, as well as the boundaries for the epicardial and endocardial surfaces.
The recording features a very fast wavefront propagating from figure-upper-left to figure-lower-right, nearly simultaneously on both surfaces (c.f. figure~\ref{fig:datafig}(c)) from a {stimulus current applied to the base of the endocardium}~\cite{gizzi2013effects}.

Over time this data displays the alternans typical of cardiac tissue excitation paced faster than the normal rhythm, namely, variation in the duration and amplitude of the action potential.
We generate observations from the dataset by constructing a cubic interpolant over space with linear interpolation over time, for both the epicardial and endocardial datasets, and sampling into this interpolant for pre-specified coordinates corresponding to the inset domain shown as dashed squares in figure~\ref{fig:datafig}(a,b).
{In the case of synthesized observations generated for the interior of the domain (cf. Section~\ref{ssec:lerp}) we linearly interpolate through the depth between the values of the observations at the projected positions on the surface to determine the value of the synthesized observation in the interior.
This ensures that we are propagating information from the surfaces into the mid-depth.}

Each $\obs{o}_i$ is likewise assigned an uncertainty $\unc_{i}$, which we identify as $\underline{\unc} = 0.05$, or $5\%$ of the action potential amplitude range, by default.
{The value of $\underline{\unc} = 0.05$ was estimated from the full-width at half-maximum of the distribution of $u^o(t) - u^o(t-\tau)$ for the dataset, with $\tau$ the assimilation interval. See figure~S2 
for a detailed motivation of this computation. }
In the case of synthetic observations interpolated into the mid-depth of the tissue (cf. Section~\ref{ssec:lerp}), we assign a larger uncertainty, $\unc_i=\overline{\unc}=0.5$, to account for the perturbative effect of these synthetic observations on the assimilated state, since the diagonal elements of the observation covariance are related to the inverse, $R_{ii} = \unc_i^{-1}$; i.e., they contain no new information from the experimental data other than a highly uncertain suggestion that the mid-plane is excited or unexcited at a particular place and time.
In the case of encoding uncertainty about the precise form of the wavefronts (cf. Section~\ref{ssec:wunc}), we perform a similar modification of the observation uncertainty, using a nonlinear function of the observation field value $\unc(\obs{o}_i)$ which is bounded by $[\underline{\unc},\overline{\unc}]$ and peaks at $\overline{\unc} = 0.5$ for $\obs{o}_i = 0.5$. 

\subsection{LETKF}

The Ensemble Transform Kalman Filter (ETKF) finds best-fit linear combinations of the prior states to match a set of observations in a quadratic sense.
The LETKF~\cite{hunt2007efficient} transforms an ensemble representing the prior covariance and mean in accordance with the local observations of the state, to affect an ensemble representing the posterior covariance and mean.
We consider the classical Kalman filter optimization functional~\cite{kalman1960new},
\begin{align}\label{eq:kalman}
    J(\ens{}) = &(\ens{}-\ensmean{b})^\top (P^b)^{-1} (\ens{}-\ensmean{b}) + \\ &(\obs{o}-H(\ens{}))^\top R^{-1} (\obs{o}-H(\ens{})), \nonumber 
\end{align}
where the prior ensemble of size $M$ is the sum of the ensemble deviation matrix ($\ensspan{b}$) and mean ($\ensmean{b}$), $\ens{b} = \ensspan{b} + \ensmean{b} [1,\dots]^\top$, similarly for the decomposition of the posterior ensemble, $\ens{a}$.
The covariance of the prior ensemble is described by the matrix $P^b = (M-1)^{-1}\ensspan{b}(\ensspan{b})^\top$.
The observations of the physical state are collected in $\obs{o} \in \mathbb{R}^{O}$, while the observation operator $H: \mathbb{R}^{N} \to \mathbb{R}^{O}$ transforms from the state space to the observation space, and matrix $R \in \mathbb{R}^{O \times O}$ describes the covariance of the observations.
The observation operator maps the state vector to the observations, with additive noise per observation with variance $\unc$, according to the uncertainty.
The minimizing solution to \eqref{eq:kalman} is denoted by the analysis ensemble, $\ensmean{a}$, which best approximates the truth state as observed by the data $\obs{o}$ and constituted from the background ensemble $\ens{b}$.

The ETKF applies the analysis update in the span of the background ensemble states.
The analysis ensemble covariance is expressed as $\tilde{P}^a = [(M-1)I/\rho + (\obsspan{b})^\top R^{-1} (\obsspan{b})]^{-1}$, where $\tilde{P}^a \in \mathbb{R}^{M\times M}$, and the analysis ensemble mean $\ensmean{a} = \ensmean{b} + \ensspan{b}\overline{w}^a$.
The scalar factor $\rho \geq 1$ is the (uniform, constant) multiplicative covariance inflation factor. 
The mean weights are determined by $\ensmean{a} = \tilde{P}^a (\obsspan{b})^{\top} R^{-1}(\obs{o} - \obsmean{b})$.
The weights of the analysis ensemble mean deviations are similarly expressed in the basis of the background mean deviations, $W^a = [(M-1)\tilde{P}^a]^{1/2}$, and $\ens{a} = \ensmean{b} + \ensspan{b}(\overline{w}^a + W^a)$.
This deviates from the error subspace transform Kalman filter (ESTKF)~\cite{nergerUnificationEnsembleSquare2012}, which performs the transformation in the ($M-1$)-dimensional subspace corresponding to the span of $\ensspan{b}$.

The LETKF extends this analysis update in the space of the background ensemble \emph{locally} -- in the sense of the observation influence and the covariance.
The inclusion of observation and covariance localization applies the preceding analysis to each element of the state vector, where the inclusion of observations for a particular state vector element is determined by the localization procedure.
In this work we use an isotropic localization weighting controlled by multiplying the background covariance matrix $P^b$ by a decaying fifth-order function, which goes to zero at a distance of $2\sqrt{10/3}\,\sigma_o$, where we have retained $\sigma_o=6$ ($21.9$ spatial units, or $\approx 0.33$~cm) for consistency with our earlier work.
We refer the reader to References~\onlinecite{hunt2007efficient, hoffman2016reconstructing, marcotte2021robust} for further mathematical details of the implementation of the localization procedure.

Finally, in lieu of an optimized multiplicative and additive inflation scheme, we used multiplicative inflation only.
The LETKF utilizes multiplicative inflation of the analysis ensemble deviations ($\ensspan{a} = \ens{a}-\ensmean{a}$) with a fixed value of $\rho=1.05$.
We did not utilize additive inflation in this work, despite previous results that showed it to be effective~\cite{marcotte2021robust}, in the expectation that our stochastic approach would similarly affect the ensemble while requiring substantially less disk space.

\section{Results}

We report results for several data-assimilation experiments. 
In all experiments, we use an observation dataset derived from optical-mapping recordings of periodic stimulation of a canine ventricle and a fixed model integration time (equivalently, assimilation interval) of $T_a = 2$~ms based on the data recording frequency.
The assimilation outputs are the background ($\ens{b}$) and analysis ensembles ($\ens{a}$), their respective means $\ensmean{*} = M^{-1}\sum_{m=1}^{M} \ens{*}_m$ where $m= 1,\dots, M$ indexes the ensemble, and their ensemble spreads,
\begin{equation}\label{eq:sprd}
     \SPRD{*} = (M-1)^{-1}\left(\sum_{m=1}^{M} |\vec{x}^{*}_m - \overline{\vec{x}}^{*}|^2\right)^{1/2},
\end{equation}   
which is the variance of the ensemble state vectors.
We have used here the vector-of-vectors notation for the ensembles, $\ens{*} = [\ens{*}_1, \dots, \ens{*}_M]$, but we may equivalently make the spatial ($ijk$), variable ($l$), and ensemble ($m$) indices explicit in a five-index array, $\ens{*} = [x^*_{ijklm}]$ where the indices $(i,j,k,l)$ are linearized for the vector form. 

We report the root-mean-square-error ($\RMSE{}$) of the reconstruction by the point-wise error of the reconstructed state by comparing the ensemble mean to the observations $\obs{o}(t)$,
\begin{equation}\label{eq:rmse}
    \RMSE{*} = \left(N_\mathrm{obs}^{-1}\sum_{i=1}^{N_\mathrm{obs}}|\obs{o}_i - \ensmean{*}_{\iota(i)}|^2\right)^{1/2}, 
\end{equation}
where $\obs{o}_i(t)$ is the $i$-th observation, and the index $\iota(i)$ maps the corresponding observation index $i$ to the corresponding element of the state vector.
Thus, the $\RMSE{*}$ summation is restricted in our experiments to the surfaces of the domain.
Depth information is available from $\SPRD{*}$; by reducing over indices of the state vector corresponding to the length and width we find a measure of the spread through the depth of the domain for each state variable,
\begin{equation}\label{eq:sprdz}
    \SPRD{*}_{kl} = \sqrt{(n_x n_y)^{-1}\sum_{i=1}^{n_x}\sum_{j=1}^{n_y} (\SPRD{*}_{ijkl})^{2}},
\end{equation}
where we have implicitly reshaped $\SPRD{*}$ from a vector to a four-dimensional array with explicit spatial ($i,j,k$) and variable ($l$) indices.
In the experiments, this scalar field is computed for the background ensemble $\ens{b}$ and $u$ field elements of the $\SPRD{b}$ over time and reported as $\SPRD{b}_u(t,z)$.

The definition of \eqref{eq:rmse} restricts its evaluation to the subset of the domain where there are observations, which tell us nothing about the consistency of the interior reconstructions.
Canonical reporting of reconstruction consistency for unknown states is handled by proper scoring rules, like the continuous ranked probability score ($\CRPS{}$) of the ensemble, which can be formulated solely in terms of the ensemble states without (necessary) recourse to observations or a `truth' state.~\cite{gneiting2007strictly}
We refer to the kernel representation of the adjusted $\CRPS{}$~\cite{leutbecher2019ensemble},
    \begin{equation}\label{eq:crps}
        \CRPS{*}(\vec{x}^b, \vec{y}) = \sum_{m=1}^{M}\frac{\|\vec{x}^b_m - \vec{y}\|}{M} - \sum_{m,{m'}=1}^M\frac{\|\vec{x}^b_m - \vec{x}^b_{m'}\|}{2M(M-1)},
    \end{equation}
with $\vec{y} = \obs{o}$ or $\vec{y} = \ens{a}$ as $\CRPS{o}$ or $\CRPS{a}$, respectively, which has desirable properties for exchangable ensembles with i.i.d. members irrespective of ensemble size $M$.
In the above, the norm ($\| \cdots \|$) corresponds to the 1-norm over the indices of the vector $\vec{y}$.
When $\vec{y}$ references the observations $\obs{o}$, then the summation is over the observations (and thus surface-restricted); when $\vec{y}$ references the analysis ensemble mean $\ensmean{a}$ the sum measures the consistency of the entire background ensemble with respect to the best approximation of the observations according to the LETKF, $\ens{a}$, over the entire state.
{We encourage the reader to peruse the Supplementary materials for a closer look at representative dynamics of the reconstruction over time.}

\subsection{Influence of model variations on reconstruction accuracy}

We have run several experiments for the reconstruction of cardiac excitation patterns.
In the simplest case, we compute the baseline accuracy of the reconstruction procedure by incorporating information about the stimulus pattern in the observations into the model dynamics, without any assimilation procedure.
Additionally, we consider a strong assimilation process without this stimulus information, to assess the importance of modeling the non-autonomous forcing inherent to the observation data for the reconstruction accuracy.
Finally, we incorporate the stimulus forcing into the model dynamics in concert with a strong assimilation procedure; this approach maximizes the information supplied to the assimilation and {provides clear measure of the importance of non-autonomous model information to the reconstruction reliability.}

\subsubsection{Free-Run  \label{ssec:freerun}}
To begin, we ask to what degree the assimilation procedure may be responsible for the reconstruction of the state, compared with the averaging procedure of the ensemble and the synchronization of the ensemble states with the stimulus cadence.
We solely assess the uncontrolled dynamics of the ensemble members with the stimulus forcing (as, in the absence of stimulus forcing, the ensemble member states all return to quiescence) for the reconstruction problem.
To this end, we report the reconstruction results with the assimilation update excised -- equivalently, the observations are not permitted to affect any of the state variables -- where the model is stimulated periodically to match the stimulus observed in the observations.

In the free-run setting we perform no assimilation by explicitly filtering all observations from the analysis computation, so that the LETKF reduces to a scaling-permutation operator as the ensemble index is not guaranteed to be preserved, and $\ensmean{a} \equiv \ensmean{b}$.
The Free-Run experiment serves as an effective method of distinguishing the improvement due to the LETKF and the stimulus model in the reconstruction over long assimilation times.

Additionally, to avoid the trivial solution, we include a stimulus current forcing in the dynamical model {for the free-run experiment}.
Thus, in this setting, we are testing the accuracy of reproducing the observations with only the model and stimulus information, without recourse to the assimilation method.
We constructed an explicit stimulus current for the ensemble model using the observation data.
We determined the period $T$, temporal offset $t_0$, and centroid position $\vec{x}_0$ for the stimulus current assuming a model of the form,
\begin{equation}\label{eq:stim}
    I_\mathrm{stim}(t,\vec{x}) = I_0 \, \Pi(t) \, \ee{-\kappa^2\left( (x-x_0)^2 + (y-y_0)^2\right)},
\end{equation}
which affects the dynamics of $u(t,\vec{x})$ by entering the evolution equations in addition to the autonomous currents $I_{so}$, $I_{fi}$, and $I_{si}$ present in the dynamical model.
The stimulus current is temporally modulated by the `top-hat' function $\Pi(t)$, which turns the stimulus current `on' for a short interval $\tau_\mathrm{on}$ ($\Pi(t) = 1$) and otherwise leaves it `off', $I_\mathrm{stim}=0$ ($\Pi(t)=0$).

The experiment which produced the dataset used in this study use a periodic biphasic stimulation, applied outside of the assimilated domain {at the base of the endocardium}.
{The stimulus current model, equation~\eqref{eq:stim}, is monophasic and independent of depth, which is not the same as the experimental stimulus current.
Rather, the our stimulus current model approximates the effect of the experimental stimulus on the observations within the assimilated domain; i.e. it is a data-driven model, and only approximates the observed effects.}
{As we only observe the effect of the experimental current some distance from its direct point of application, we effectively filter the experimental stimulus current through the tissue observations to construct our stimulus current model}.
Throughout, we set $I_0 = 0.1$~ms$^{-1}$, with the duration of the stimulus set to $\tau_\mathrm{on} = 10$~ms, and set the spatial scale of the stimulus current to $(\kappa/h_x)^2 = 5\times 10^{-4}$ to match that observed in the {observation data}.
The {period } between successive stimuli is $118$~ms, likewise matching the observation data.
{Tests were performed to assess the impact of the stimulus current parameters on the state, which were found to tall into two classes: (a) sub-critical (non-excitatory) and (b) super-critical (excitatory) stimuli, which showed no meaningful variation for physically realistic values within each class. This critical excitation phenomenon is well-understood in the one-dimensional case.~\cite{idris2007critical,simitev2006conditions,marcotte2020predicting}}

\begin{figure}[htpb]
    \centering
    \includegraphics{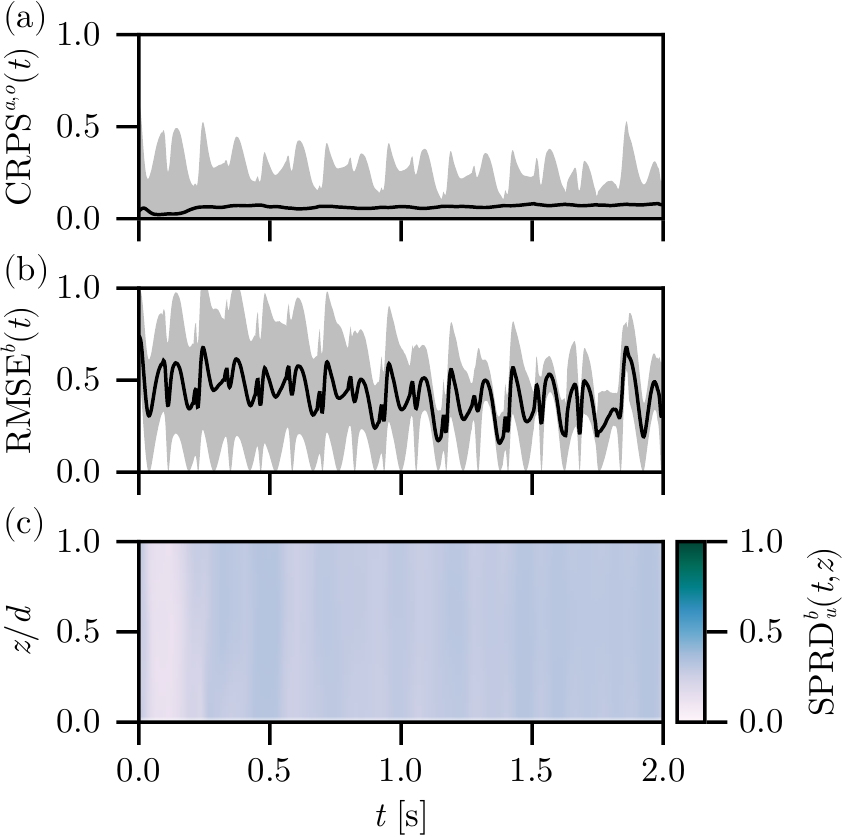}
    \caption{Free-Run experiment results, depicting the (a) $\CRPS{a}(t)$ (line) and $\CRPS{o}(t)$ (band), (b) $\RMSE{b}(t)$ (line) plus and minus one standard deviation (band), and (c) $\SPRD{b}(t,z)$ (color) for the reconstruction using the Barone \emph{et al}. parameter set.}
    \label{fig:freerun}
\end{figure}

In figure~\ref{fig:freerun}, we show the $\CRPS{a,o}(t)$, the $\RMSE{b}(t)$, and the spread through the depth of the tissue $\SPRD{b}_u(t,z)$ for the Free-Run experiment using the Barone \emph{et al}. parameter set.
The $\CRPS{o}(t)$ and $\CRPS{a}(t)$ indicate signficant disagreement between the background ensemble and the observations, as well as in comparison to the analysis ensemble mean.
The $\RMSE{b}(t)$ indicates that the ensemble does not accurately reconstruct the surface dynamics in the absence of assimilation, as expected.
Furthermore, the variability of $\RMSE{b}(t)$ decreases over time, as the forcing of the stimulus current synchronizes the large-scale features of the ensemble states as information about the ensemble initialization is slowly lost.
Likewise, the $\SPRD{b}_u(t,z)$ is nearly invariant with respect to $z$, indicating that we correctly restore the $z$-invariance symmetry of the ensemble states when the observations are not present in the reconstruction procedure.

These results serve as a baseline for comparing with the experiments to follow. 
For the Free-Run experiment we have used our unmodified observation set and standard ensemble initialization, with the stimulus model and absent any assimilation.
The ensemble means therefore do not correspond to a reconstruction, as there are no guiding observations, and so we do not show the mean fields.

\subsubsection{Autonomous model \label{ssec:strong}}
The first experiment uses an autonomous, deterministic, cardiac excitation model for the ensemble members in the assimilation of these observations to consider the significance of model error arising from model parameters which may or may not effectively reproduce the observed action potentials.
Instead of adapting the model to every stimulus protocol, it is dramatically simpler to use the properties of the Kalman filter to match the periodicity of the stimulation without explicit recourse to a modelled stimulus pattern. 
This approach has the obvious limitation that we have rejected a potential source of additional information and introduced latency into the assimilation, as the ensemble can only react to the next stimulus \emph{after} it has already appeared in the observations.
Given this restriction, this experiment shows the worst-case, `low-information', estimate of the state of the system using the LETKF assimilation infrastructure. 
We refer to this experiment as `Autonomous'.

\begin{figure}[htpb]
    \centering
    \includegraphics{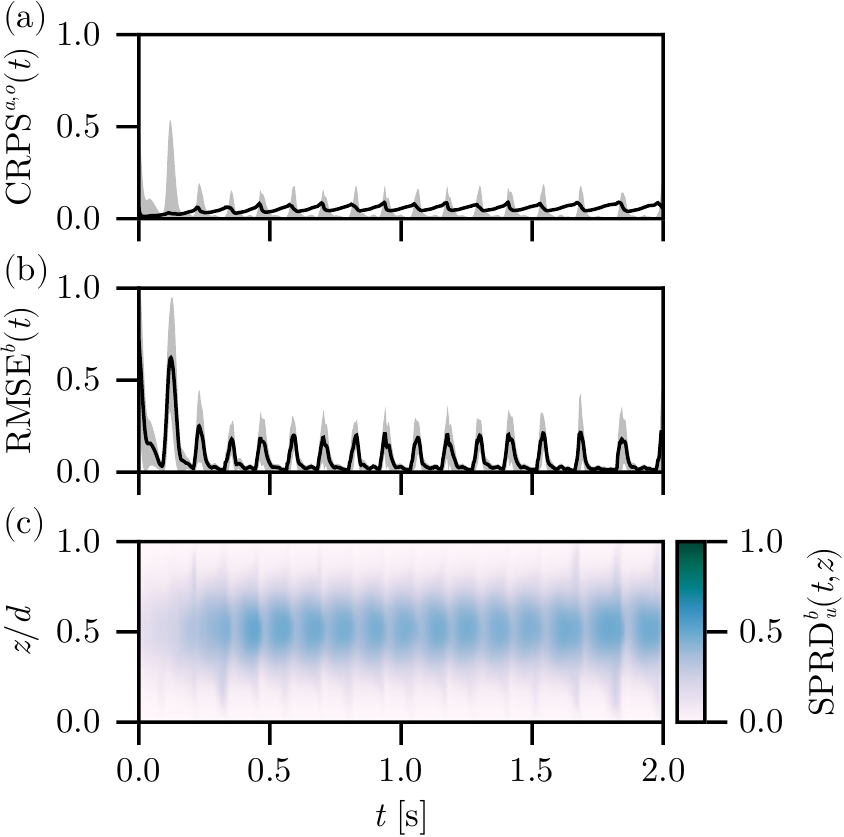}
    \caption{Autonomous experiment results, depicting the (a) $\CRPS{a}(t)$ (line) and $\CRPS{o}(t)$ (band), (b) $\RMSE{b}(t)$ (line) plus and minus one standard deviation (band), and (c) $\SPRD{b}(t,z)$ (color) for the reconstruction using the Barone \emph{et al}. parameter set.}
    \label{fig:auto}
\end{figure}

Figure~\ref{fig:auto} shows the (a) $\CRPS{a,o}(t)$, (b) $\RMSE{b}(t)$, and (c) $\SPRD{b}_{u}(t,z)$ metrics over time for the reconstruction using the Fenton-Karma model with the Barone \emph{et al.} parameter set.
The assimilation does a very good job of reconstructing the surfaces of the state to match the observations after the initial transient, which arises from the phase mismatch of the initial ensemble and the observations, producing high initial error.
The $\CRPS{o}(t)$ is strongly correlated with the $\RMSE{b}(t)$, as expected, and $\CRPS{a}(t)$ remains small over long time scales, growing slowly and oscillating due to the underlying excitation period.
This behavior is due to the lag inherent to this assimilation experiment, due to the information of the next excitation needing to propagate through the full LETKF before affecting the background ensemble.
The dominant contribution to the $\RMSE{b}(t)$ corresponds to a peak occuring during the full excitation of the domain, with fast relaxation of the error as the excitation amplitude dissipates.
One interpretation of this contribution is that, on the surfaces, the LETKF struggles with the uncertainty of the wavefront and deviation of the recorded wavefront shape from those generated by the nonlinear model for the background ensemble (the model lags the observed wavefront, by construction), but manages to reproduce the waveback accurately.
Indeed, the innovation of the LETKF is large during the propagation of the wavefront---the LETKF is correcting a slow CV in the model to match the excitation pattern observed in the data---and relatively smaller during the propagation of the waveback.
The spread for this parameter set begins small and quickly grows in the interior, representing the dominant contribution of the uncertainty of the state of the system, as expected.
As the information about the excitation propagates from the surfaces toward the interior of the domain, it is clear that the LETKF is unable to constrain the interior of the solution using only the surface observations.
That said, the maximum of the spread at $z/d=1/2$ saturates after $t\approx250$~ms, indicating that this time period reflects the true uncertainty of the interior.

On the surfaces, after an excitation, the next action potential (AP) is driven by the assimilation perturbing the state in an appropriate region to cause a new propagating wave that matches those in the observations, as expected.
Additionally, as the different parameter sets generate APs with different wave speeds ( $c_\parallel \approx \sqrt{D_\parallel/\tau_d}$), the assimilation continually corrects the propagation directly ahead of the wavefront. 
When the wave is slightly too slow, this correction is effective as the region preceeding the wavefront is expected to be excitable;  thus small perturbations ahead of the wave create a new front that connects to the existing lagged wavefront before the next assimilation time.
When the wave is slightly too fast, this correction is ineffective, as reducing the value of $u$ by a small amount on the wavefront does not significantly hinder the propagation---doing so requires a larger, quenching, perturbation related to the gap between the middle unstable nullcline of $I_{fi}$ and the rest state.
This asymmetry in effective control is apparent on the surfaces because of the presence of observations, and likewise means uncontrolled wave speed errors propagate in the interior. 

\begin{figure}[htpb]
    \centering
    \includegraphics{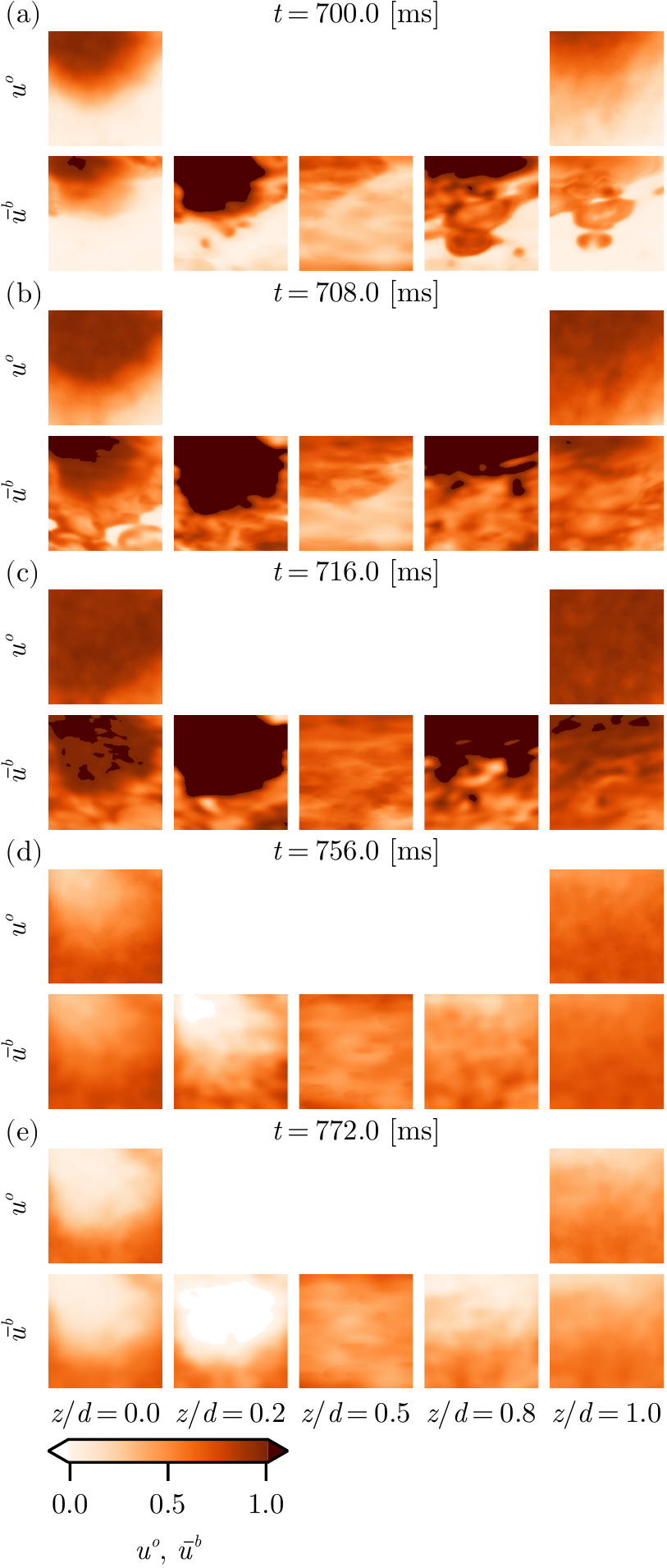}
    \caption{Slices of the the $u$-field of the observations ($u^o$) and the background ensemble mean ($\overline{u}^{b}$) at depths $z/d = 0.0,\, 0.2,\, 0.5,\, 0.8,\, 1.0$ and times $t=700,\,708,\,716,\,756,\,772$~ms for (a-e), respectively.}
    \label{fig:assimslices}
\end{figure}

The phenomenology of the assimilation dynamics is exemplified by figure~\ref{fig:assimslices}, which shows $z$-slices of the background ensemble mean $\ensmean{b}$ at fixed depths for three sample times from the Autonomous experiment.
Near the beginning of a new excitation in the observations, cf. figure~\ref{fig:assimslices}(a), the ensemble mean is a very good approximation of the distribution of the excitation at the surfaces ($z/d=0.0,1.0$) while the interior slices are not reasonable approximations to an interpolating function through the depth.
Rather, the interior is too uncertain to match the propagation of a wave across the domain, which we expect to pass uniformly in $z$, in time with the surfaces.

Further, these slices highlight a significant contribution to the model error: the timing of the front in the ensemble is lagged compared to the observation data.
This behavior is due to the interaction of three phenomena related to the timing of the periodicity of the observations and how these synchronize the ensemble dynamics.

The first and most significant is the slower conduction velocity of the model than the observation data.
{Despite being manually tuned to the data~\cite{barone2020experimental}, the parameter set is unable to adequately match the observed front speed without a significant overestimate of the tissue conductivity (cf. Ref.~\onlinecite{barone2020experimental}, their $\sigma_{\parallel}=9\times D_{\parallel}$ and $\sigma_{\perp}=15\times D_{\perp}$)~\footnote{In our experiments these values yield faster wave propagation, but also strongly drive the ensemble states to quiescence, dramatically worsening the overall reconstruction with (trivial) ensemble collapse. Even when including the stimulating current into the model dynamics, the reconstruction error is significantly worse (not shown). Hence, we use more physiologically reasonable values for the tissue conductivity in this work.} and thus any timing of the excitation wave in the ensemble with the observations becomes a significant, ongoing, correction for the LETKF. }
{Previous \emph{in silico} studies have found the LETKF to adequately address a variety of systemic model errors in 3D, including fiber angle rotation, tissue conductivity, and the explicit time scales of the model.~\cite{hoffman2016reconstructing, hoffman2020sensitivity, marcotte2021robust}}
Ahead of the wavefront, the LETKF can readily excite the leading edge of quiescent tissue in the ensemble member states, speeding up the front to match the excitation propagation in the observations, provided there are ensemble members that are already excited in that leading region.
Due to the slower conduction velocity, this scenario is rarely the case, as the ensemble members synchronize their phase with the observations due to the driving of the LETKF.

The second contributor to the model error is the lag inherent to the propagation into the interior of the domain for the autonomous model.
The interior dynamics are lagged by at least $\tau \approx c_\perp d/2$, the time it takes information on the surfaces to propagate to the interior for a transverse wave speed of $c_\perp$. 
Thus, even when the surfaces correctly estimate the distribution of the excitation in the observations, the interior distribution is dominated by the dynamics from several assimilation steps ago.
These discrepancies are systematic errors in the reconstruction that are largely invisible to the $\RMSE{b}$ metric, as it relies on the surface observations, but should affect $\CRPS{a}$ if the ensemble members differ in the interior, and likewise $\SPRD{b}_u(t,z)$.

The third contributor is the lag associated with the autonomous model dynamics due to the lack of stimulus information.
During quiescence, the next excitation wave is observed at time $t$; these observations are used in the LETKF and their influence appears in the analysis ensemble states at time $t$.
These analysis ensemble states form new initial conditions for the dynamical model, and the effect of this perturbation due to the initial observations appears in the next background ensemble, at $t+T_a$, thereby
introducing a lag between the background ensemble and the observations of no less than the assimilation interval time.
If the threshold for excitation is not met for the analysis ensemble state, then the lag between the observed excitation stimulus and the resulting wave will be larger, which is typically the case as the initial observations of an excitation are, by definition, small, and likely sub-threshold.

Finally, we must note several features of the observations and reconstruction that will become relevant in later results.
The observations smoothly interpolate from quiescent ($u(t,x,y,z) \approx 0$) to fully excited ($u(t,x,y,z) \approx 1$) across the wavefront, cf. figure~\ref{fig:assimslices}(a).
Such blunted fronts represent a configuration not achievable with the model at hand---this model exhibits a threshold, such that the wavefront is sharply defined; i.e., $u(t,x,y,z)$ is very likely to be close to $0$ or close to $1$, and the probability that $u(t,x,y,z)$ is near $0.5$ along the wavefront is exceptionally small.
The observations suggest that, at this particular time, $u^o$ has a signficant probability of being close to $0.5$.
This tension between the spatially smooth data and the sharp features of the model forms a significant impediment to reliable reconstruction.

Additionally, the dynamics of the model with the chosen parameter set---despite being manually tuned for this dataset---tends to overshoot the action potential amplitude (APA) of the observations, c.f. figure~\ref{fig:assimslices}(a-c).
This behavior is caused, first, by the analysis step of the LETKF over-correcting the ensemble states.
The effect then persists due to the relatively short model integration interval ($T_a = 2$~ms), which is not sufficient for the state to relax onto the slow manifold.
This effect may also be observed in the Free-Run experiment due to the relatively high frequency of the stimulus current (not shown).
Notably, the effect of the high-frequency pacing stimulus current and the overshoot of the analysis step are not additive.

\subsubsection{Explicit stimulus current modeling \label{ssec:stim}}

This experiment considers the effect of including a non-autonomous, spatially localized, explicit stimulation current in the dynamics of the model that is tuned both spatially and temporally to qualitatively match the stimulation observed in the experimental dataset, and the potential for catastrophic synchronization of the ensemble in the presence of this explicit modeling choice.
The stimulus current used is the same as that used in the Free-Run results (Section~\ref{ssec:freerun}).
We refer to this experiment as `Stimulus'.

\begin{figure}
    \centering
    \includegraphics{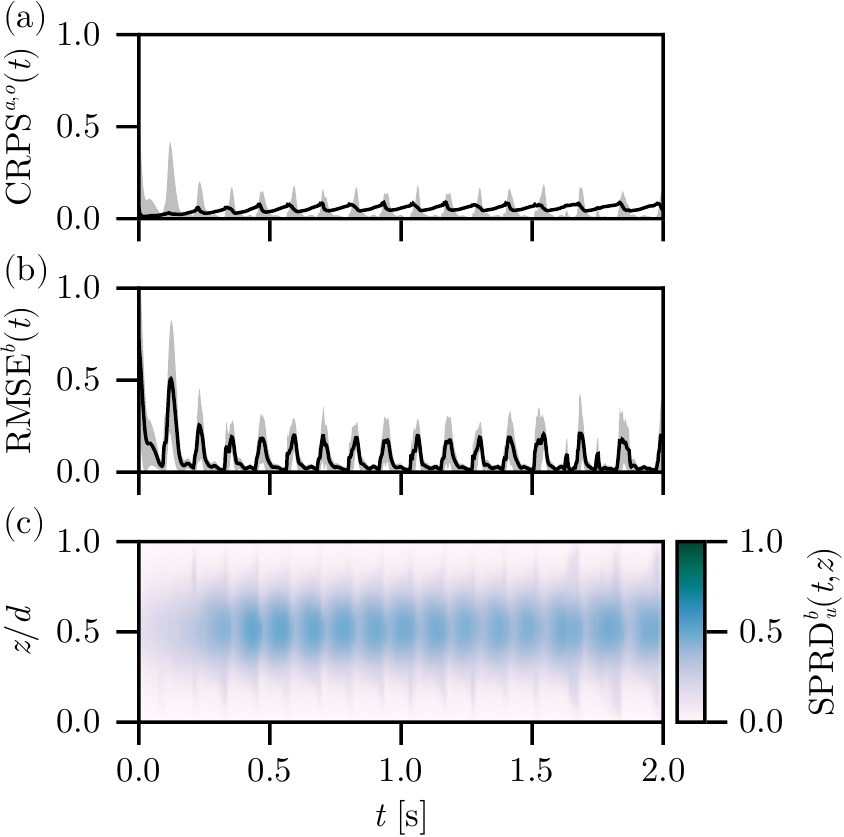}
    \caption{Stimulus experiment results, depicting the (a) $\CRPS{a}(t)$ (line) and $\CRPS{o}(t)$ (band), (b) $\RMSE{b}(t)$ (line) plus and minus one standard deviation (band), and (c) $\SPRD{b}(t,z)$ (color) for the reconstruction using the Barone \emph{et al}. parameter set.}
    \label{fig:stim}
\end{figure}

The results for the model are shown in figure~\ref{fig:stim}, where the reconstruction is largely insensitive to the inclusion of the stimulus current.
Several small changes appear in the $\RMSE{b}(t)$ compared to the Autonomous results.
The reconstruction error exhibits smaller peaks during the excitation phase, where the stimulus current triggers the excitation wave ahead of the lag associated with the observation-LETKF-model path in the Autonomous experiment.
After $t = 1.5~\mathrm{s}$, drift in the timing of the stimuli leads to an early stimulus, which is corrected by the LETKF on the surfaces, effectively constraining the timing error.
These early stimuli appear as small ($\lesssim 0.1$) increases in the $\RMSE{b}(t)$ from the baseline outside of the dominant contributions from the excitation wave in figure~\ref{fig:stim}(a)---the short-lived effect on the surface error is also an indication that the reconstruction procedure is robust with respect to errors in the specification of the stimulus period, a prime concern due to the timing-related contributions covered in the Autonomous reconstruction.

\begin{figure}[htpb]
    \centering
    \includegraphics{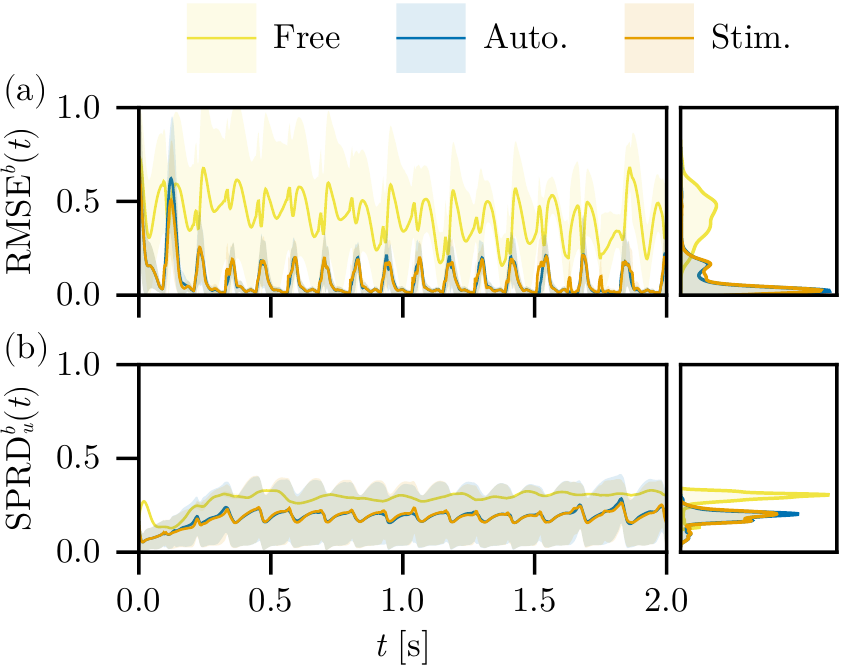}
    \caption{(a) $\RMSE{b}(t)$ and (b) depth-average $\SPRD{b}_u(t)$ for the Free-Run, Autonomous, and Stimulus experiments, with the distribution of (right, top) surface errors and (right, bottom) ensemble spread over time.}
    \label{fig:autostimfreemarginal}
\end{figure}

In figure~\ref{fig:autostimfreemarginal} we show the (a) $\RMSE{b}(t)$ for the Autonomous, Stimulus, and Free-Run experiments over time (left), and the marginal density plots of the surface errors (right), for the reconstruction using Fenton-Karma and the Barone \emph{et al}. parameter set.
The improvement of the $\RMSE{b}$ over the Free-Run experiment is significant, but the surface errors for the Autonomous and Stimulus experiments are comparable, both overall and over time.
The most significant correction of the $\RMSE{b}(t)$ for the Stimulus results over the Autonomous model is a shorter initial transient, as the first excitation ($t \approx 125$~ms) reaches a lower amplitude error on the surfaces, and thus the ensemble synchronizes to the true stimulus rhythm more quickly.
The $\RMSE{b}$ makes it clear that the inclusion of the LETKF effectively lowers the expectation value of the surface error, overall.
For both the Autonomous and Stimulus experiments, the surface error is nearly identical -- it appears that the inclusion of stimulus information into the model dynamics has a minimal impact on the surface reconstruction error over long times, save for the minor advantages noted above.
The depth-averaged $\SPRD{b}_u(t)$ for the Free-Run experiment is larger than in either the Autonomous experiment or Stimulus, reflecting the low-confidence of the reconstruction in the absence of the LETKF.
Indeed the depth-averaged $\SPRD{b}_u(t)$ is nearly identical for the Autonomous and Stimulus results, with a subtle bias in the latter toward larger variation over time, cf. figure~\ref{fig:autostimfreemarginal}(b).

\begin{figure}[htpb]
    \centering
    \includegraphics{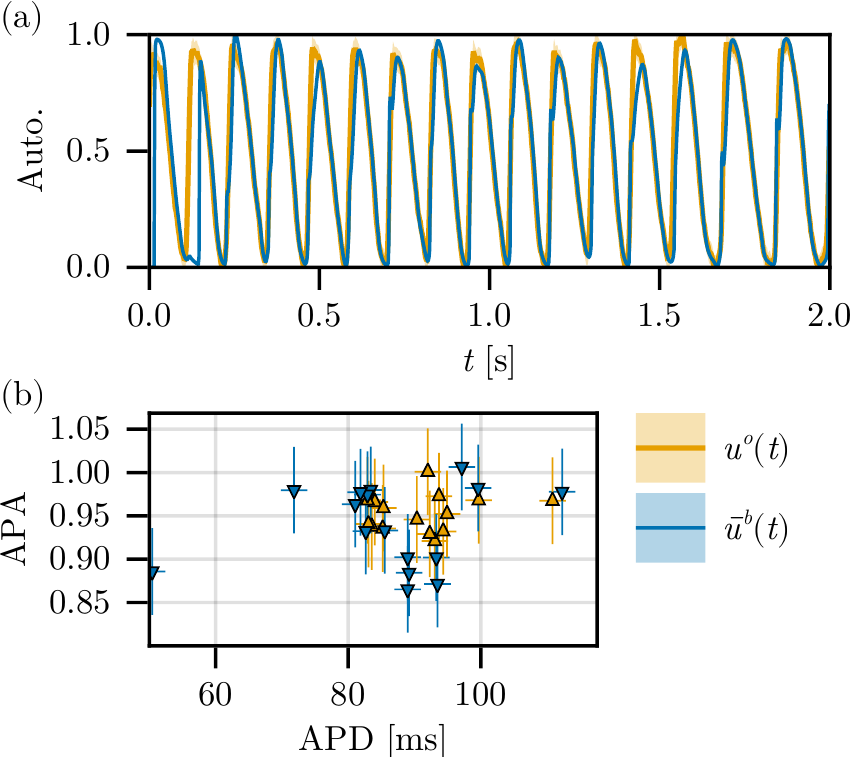}
    \caption{{(a) Temporal trace for the Autonomous experiment and associated observations, sampled on the epicardium ($z/d=0$) at $(x,y)=(0.825,0.42)$~cm, and (b) corresponding action potential durations (APD) and amplitudes (APA) ($u_\mathrm{thr} = 0.1$).}}
    \label{fig:trace_APD}
\end{figure}

{In addition to the reconstruction error of the field, it is worth considering the error in physiologically relevant quantities which are derived from the state, e.g. those which inherit from the structure of the action potential, including the action potential duration (APD) and amplitude (APA).
We report the temporal trace of the autonomous reconstruction for a randomly selected position in the observations in figure~\ref{fig:trace_APD}(a).
After an initial transient which lasts less than two BCL ($<250$~ms) the reconstruction at this point is excellent, with minor errors appearing in the apex of the action potential, and thus contributing to a poor estimate of the true APA, cf. figure~\ref{fig:trace_APD}(b).
The distribution of APD between the observation and reconstruction traces match exceptionally well, as expected because the predominant error in the reconstruction appears far from the threshold value ($u_\mathrm{thr} = 0.1$).
Similar calculations for all LETKF-driven reconstructions are included in the Supplement, cf. Figure~S1.} 

\subsubsection{Stochastic current model effects \label{ssec:stoch}}

This experiment incorporates stochasticity into the model to investigate the efficacy on the accuracy of state reconstruction in the context of uncontrolled model errors.
We have previously introduced these stochastic model effects for the assimilation of observations generated from state sequences generated by a model, subject to model uncertainty~\cite{marcotte2021robust}.
These techniques improved the reconstruction of the state over LETKF in concert with a deterministic ensemble model, occasionally by significant margins; their introduction for assimilating experimental observations may help to recover the state dynamics when the model error is due to the model being unable to perfectly reproduce experimental data.
For this work we will introduce a stochastic current (SDE-$u$) and stochastically selected time-scale parameters (SMP-$\tau$) into the model.
The SDE-$u$ formulation introduces a new stochastic current $I_\mathrm{sto}$ in the role of $I_\mathrm{stim}$ in the dynamics of the voltage variable $u$ in  \eqref{eq:fk},
\begin{equation}\label{eq:sde}
    I_\mathrm{sto} = \xi, \quad \xi \sim N(0,\sigma_u^2),
\end{equation}
while the SMP-$\tau$ affects the selection of all the explicit time-scales of the model,
\begin{equation}\label{eq:smp}
    \tau_* \sim N(\overline{\tau}_*, (\overline{\tau}_*\sigma_p)^2),
\end{equation}
where $\overline{\tau}_*$ is the baseline parameter value.
We refer to this experiment as `Stochastic'.
We consider fixed standard deviations ($\sigma_u = 0.04$ and $\sigma_p=0.04$) for both stochastic effects.

\begin{figure}[htpb]
    \centering
    \includegraphics{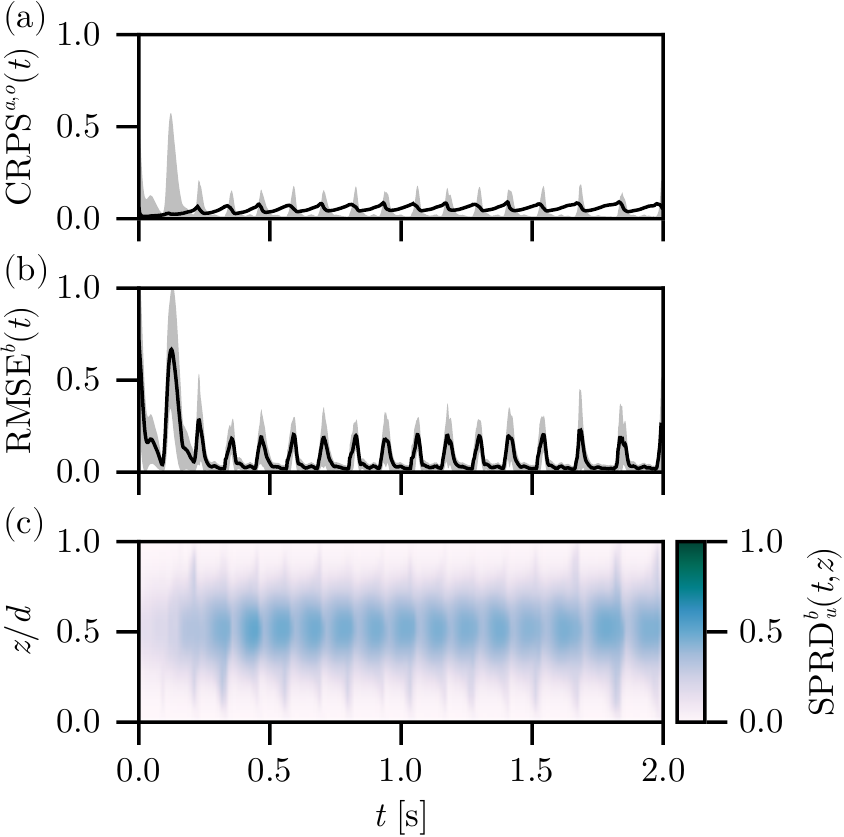}
    \caption{Stochastic experiment results, depicting the (a) $\CRPS{a}(t)$ (line) and $\CRPS{o}(t)$ (band), (b) $\RMSE{b}(t)$ (line) plus and minus one standard deviation (band), and (c) $\SPRD{b}(t,z)$ (color) for the reconstruction using the Barone \emph{et al}. parameter set.}
    \label{fig:stoch}
\end{figure}

In figure~\ref{fig:stoch}(a) we show the surface errors $\RMSE{b}(t)$ of the reconstruction for the Fenton-Karma model with Barone \emph{et al}. parameter set, with the current-based SDE and SMP-$\tau$ stochastic effects.
For these relatively low-amplitude stochastic effects, we find very minor changes to the surface reconstruction error---an effective smoothing of the peaks of $\RMSE{b}(t)$ corresponding to the excitation wavefront and waveback, which is expected from a uniformly stochastic effect.
Analogously with the Stimulus result, the transient in the $\RMSE{b}(t)$  is slightly worsened by the inclusion of the Stochastic effects; the excitation near $t\approx 125$~ms exhibits a larger peak error.

In previous work, we showed that the stochastic additions were capable of permitting the ensemble to better approximate the local degrees of freedom of the true dynamics by effectively inflating the spread for a given ensemble size.
In this experiment, we find no such effect.
We find that the addition of the stochastic current accelerates the depth-averaged $\SPRD{b}_u(t)$, leading to faster average spread growth over the initial phase, before switching to a conservative effect over the long-term reconstruction, suppressing short-lived variations in the spread.
Indeed, what this behavior indicates is that the addition of stochastic processes to the model may \emph{enhance} synchronization of the ensemble in some scenarios, lessening the ensemble spread, or \emph{suppress} total decoherence for the ensemble.

Comparing the surface errors $\RMSE{b}(t)$ corresponding to the autonomous (figure~\ref{fig:auto}), stimulus (figure~\ref{fig:stim}), and stochastic (figure~\ref{fig:stoch}) model experiments, we find that the inclusion of the stimulus current or stochastic effects has a small effect on the overall reconstruction error. 
We may interpret this as, at best, a marginal improvement in the accuracy of the surface reconstruction with the inclusion of the stochastic or stimulus current into our dynamical model.
Alternatively, we may interpret the autonomous result as achieving reconstruction errors lower than expected and with confidence comparable to these more informed model approaches.
Likewise, due to the slow drift of the timing of the stimulus current in the observations and model, we may be confident in the robustness of the reconstruction with respect to model errors related to stimulus timing.

\subsection{Influence of observation perturbations on reconstruction accuracy}

If modifications to the model have only subtle effects on the reconstruction of the state dynamics, then perhaps the tuning of experimental observations will have more significant effect.
In these experiments, we perturb the observations from the experimental data to better model the uncertainty associated with the physical processes involved in the recordings, or to buttress the highly-sparse observations with some intermediate information based on the apparent temporal structure of the data.
Likewise, this investigation presents an opportunity to ensure that our reconstruction is robust for the type of data available to the cardiac researcher.

\subsubsection{Addition of synthetic observations \label{ssec:lerp}}

In this experiment, we explore the generation of mid-depth synthetic observations, in addition to those on the surfaces, for use in the assimilation process.
The role of these synthesized observations is to gently constrain the ensemble spread in the interior of the tissue, which we have previously identified as a predominant source of error in the reconstruction of the state in this system~\cite{marcotte2021robust}.
{The synthesized interior observations are interpolated between the observation values from their projections onto the epicardial and endocardial surfaces. }
The synthetic observations are appended to the real observations, the former having fixed point-wise uncertainty estimates of $\unc=0.50$ and the latter having fixed point-wise uncertainty estimates of $\unc=0.05$.
Different spatial distributions of mid-depth observations can have a significant effect on the robustness and accuracy of the reconstruction.
A full arrangement of observations that sample the $(x,y)$-grid in the same positions as on the surfaces led to immediate ensemble collapse due to strong contraction of the analysis step (not shown).
We use a relatively sparse arrangement where $64$ observations were distributed uniformly in an $8\times 8$ grid, {which } did not collapse the ensemble.

We add these synthesized observations due to the spatiotemporal pattern of the underlying excitations, where the wave on the surfaces passes nearly uniformly over $(x,y)$ and in time.
{We thus treat the observations on each surface as strongly coupled and thus amenable to interpolation through the depth, i.e., between the surface sets. }
We report the impact the $64$ additional observations (a $0.7\%$ increase in the total number of observations) has on the reconstruction fidelity.

\begin{figure}
    \centering
    \includegraphics{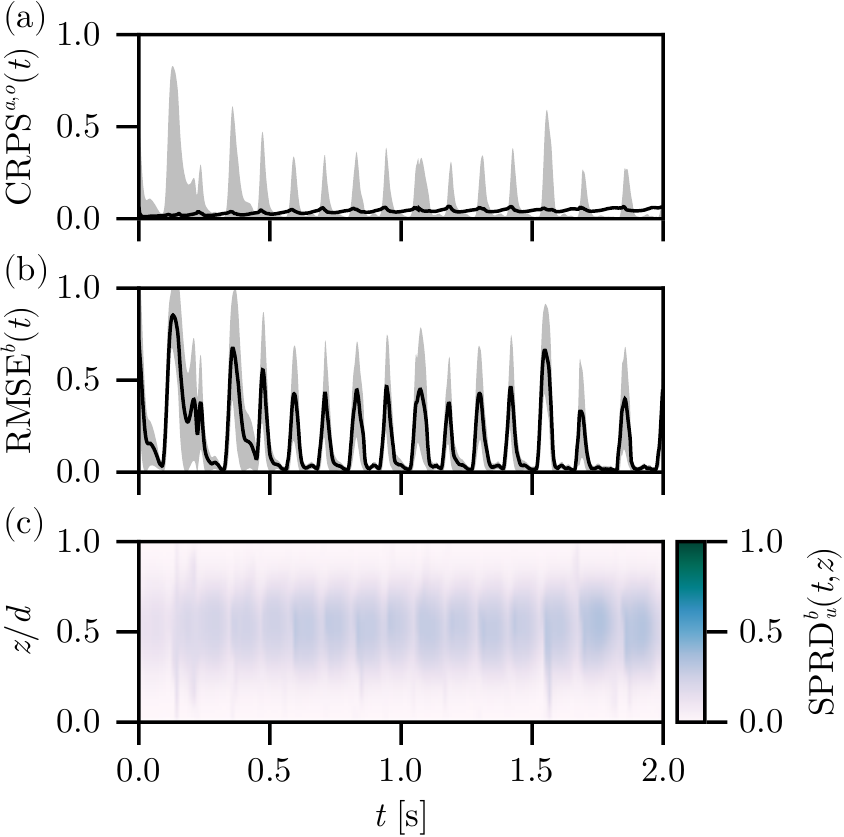}
    \caption{Synthetic Observations experiment results, depicting the (a) $\CRPS{a}(t)$ (line) and $\CRPS{o}(t)$ (band), (b) $\RMSE{b}(t)$ (line) plus and minus one standard deviation (band), and (c) $\SPRD{b}(t,z)$ (color) for the reconstruction using the Barone \emph{et al}. parameter set.}
    \label{fig:lerp}
\end{figure}

The results of the assimilation with the synthetic mid-depth observations are depicted in figure~\ref{fig:lerp}.
Immediately, we note that while $\CRPS{a}(t)$ is not significantly affected, $\CRPS{o}(t)$ now peaks significantly higher than any other assimilation experiment, cf. figure~\ref{fig:lerp}(a).
This finding immediately indicates a significant deviation from the observations, but does not reveal whether this discrepancy is due to the interior, synthesized observations, or the `true' surface observations.
The contribution is distinguished by $\RMSE{b}(t)$, which likewise peaks significantly higher than any other assimilation experiment; cf. figure~\ref{fig:lerp}(b), and is computed over the surface observations only for consistency with the other reconstruction experiments.
Notably, the $\RMSE{b}(t)$ may be decomposed into contributions from the surfaces and the synthetic observations in the interior; doing so would provide a direct estimate of the error of the reconstruction in the interior provided our interpolation ansatz is correct and reasonable.
However, we have no rigorous way to assess the interpolation ansatz and so cannot assert the correctness of the reconstruction in the interior using this method.
So we conclude that the inclusion of the interior synthetic observations interpolated from the surface observations leads to significantly worse reconstructions on the surfaces, despite the high uncertainty of the synthetic observations.

The resolution of this puzzle may appear from a consideration of the ensemble spread.
While additional mid-depth observations are extremely good at restricting the uncertainty of the reconstruction in the mid-depth of the domain, cf. figure~\ref{fig:lerp}(c), as information from the surface observations can affect the interior layers of the solution, this effect is not always desirable.
By constraining the interior, we limit the spread of states used in the reconstruction of the surfaces, as well.
This now-restricted ensemble of states may not adequately cover the physical instabilities present in the observed surface dynamics, limiting the accuracy of the reconstruction for observations for which we are confident (those on the surface) for the sake of observations for which we are not (those in the interior).
Tuning the uncertainty and distribution of the interior observations compared to the surface observations to critically constrain the ensemble thus presents an additional window for optimization, although it is beyond the scope of this work.

\subsubsection{Observation uncertainty modeling for fronts  \label{ssec:wunc}}

For this experiment we seek to infuse the observational data with uncertainties based on our knowledge of the physical and numerical filtering of the experimental data and the dynamics of the model.
To this end, we identify the slow dynamics of $u$ near the rest state ($u\approx 0$) and in the excited state ($u \approx 1$), and note that while the model produces sharp wavefront features, the observation data is relatively smooth.
In practice, this assigns an observation uncertainty that depends on the observation value, $\unc_i = \unc(\obs{o}_i)$, and specifically on the value of $\underline{u} < u < \overline{u}$, where $\underline{u} = 0.1$ corresponds to a threshold marking the boundary of quiescence and $\overline{u} = 1.0-\underline{u}$ marks the boundary of the fully excited state.
The nonlinear uncertainty function takes the form of a scaled and truncated Gaussian, 
\begin{equation}
    \unc(\obs{o}_i) = \underline{\unc} +(\overline{\unc}-\underline{\unc}) \exp\left(-k^2(u_i - (\overline{u}+\underline{u})/2.0)^2\right),
\end{equation}
where $\underline{\unc} = \underline{u}/2 = 0.05$ and $\overline{\unc} = 0.50$ are lower and upper bounds for the uncertainty, and $k$ is chosen so that $\unc(0.0) = \unc(1.0) = \underline{\unc}$ and $\unc((\overline{u}+\underline{u})/2) = \overline{\unc}$.
The simplest interpretation of the wavefront uncertainty maximum $\overline{\unc}$ is that anywhere along the wavefront, the smoothing of the observation data makes it unclear whether an ensemble member should be in a fully excited or quiescent state at the same point in time, i.e., $|\overline{u}^b - \obs{o}_i| \sim U(-\overline{\unc},+\overline{\unc})$.
In principle, this approach encourages the growth of the ensemble spread along the wavefront, so that the ensemble mean is a good approximation to the observation data while the individual ensemble members retain their sharp wavefront features.
Tuning the precise distribution of uncertainty, whether through a different ($\overline{\unc}$, $\underline{\unc}$) pairing or differently motivated nonlinear function $\unc(\obs{o})$ altogether, is beyond the scope of this work; in the present investigation we seek to perturb the existing observation data to determine if {the reconstruction accuracy may be improved. }

\begin{figure}[htpb]
    \centering
    \includegraphics{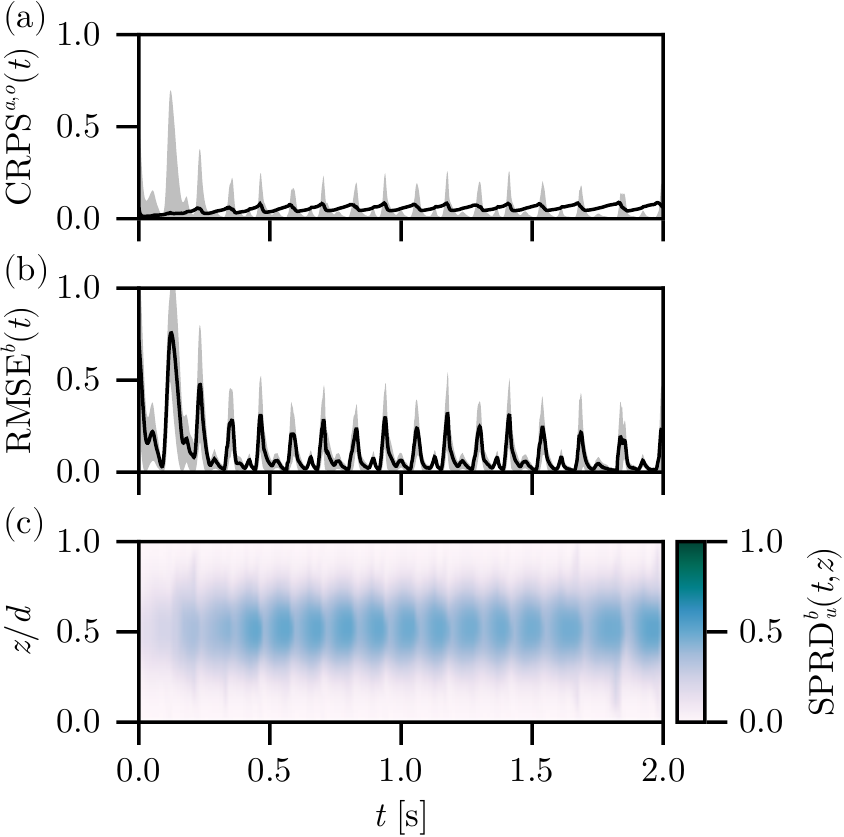}
    \caption{Uncertain wavefront experiment results, depicting the (a) $\CRPS{a}(t)$ (line) and $\CRPS{o}(t)$ (band), (b) $\RMSE{b}(t)$ (line) plus and minus one standard deviation (band), and (c) $\SPRD{b}(t,z)$ (color) for the reconstruction using the Barone \emph{et al}. parameter set.}
    \label{fig:wunc}
\end{figure}

In figure~\ref{fig:wunc}, we report the effects of the Wavefront Uncertainty experiment. 
We find that the additional uncertainty along the wavefronts and wavebacks in the observation data leads to a small, but consistent, increase in the $\RMSE{b}(t)$ near the end of the wavefront, and likewise increases the surface error during the end of the waveback, introducing a set of smaller peaks in $\RMSE{b}(t)$ between the usual set associated with the end of the wavefront.
As we are encouraging deviations in the ensemble members near the beginning and end of the excitation wave, this finding is not unexpected---exactly in the transition between excited and unexcited, the spread is increased slightly, and it makes the most significant relative contribution to the surface error where and when the error is already low.
Additionally, as the observations are synchronizing forces for the ensemble subject to the inverse observation covariance weighting, $R^{-1}$, a higher uncertainty for the observations should lead to less synchronization overall and higher ensemble spread.
We observe faster initial growth of the $\SPRD{b}_u(t,z)$ compared to the Autonomous experiment, but the long-term maximum is not significantly altered. 

\begin{figure}[htpb]
    \centering
    \includegraphics{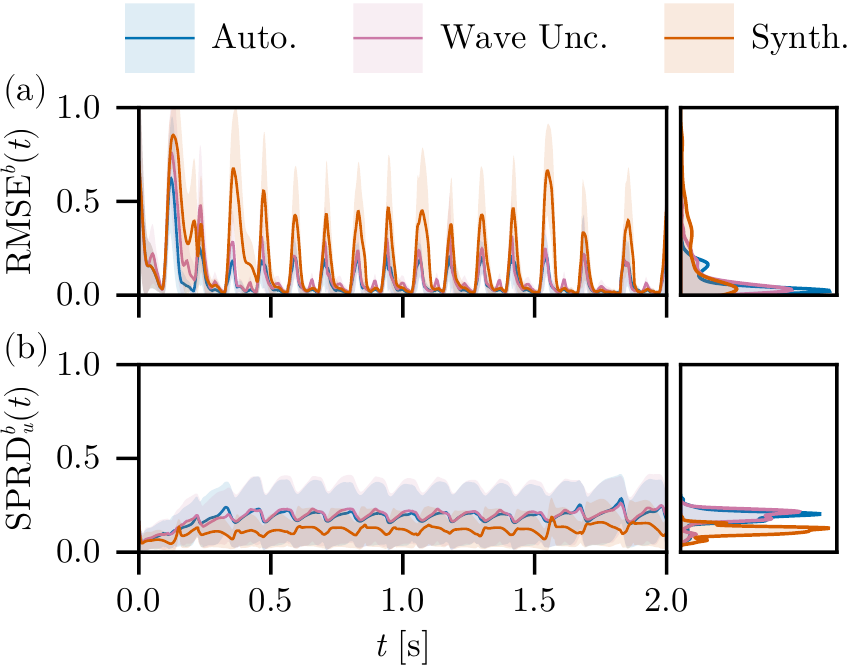}
    \caption{(a) $\RMSE{b}(t)$ and (b) $\SPRD{b}_u(t)$ for the Autonomous, Wave Uncertainty, and Synthetic observations experiments, with the distribution of (right, top) surface errors and (right, bottom) ensemble spread over time.}
    \label{fig:obserrs}
\end{figure}

In figure~\ref{fig:obserrs}, we show the $\RMSE{b}(t)$ and $\SPRD{b}_u(t)$ for the Autonomous, Wavefront Uncertainty, and Synthesized Observations experiments.
Both the wavefront uncertainty and synthesized observations experiments lead to larger surface reconstruction errors ($\RMSE{b}$) overall compared to the Autonomous experiment, though their relative contribution over time is highly non-uniform.
Whereas the Autonomous experiment reconstruction manages to match the APD and DI of the observation data, the Synthesized Observation experiment reconstruction undergoes large deviations from accurate reconstructions, frequently lasting longer than the APD of the observations.
Likewise, while the Wavefront Uncertainty experiment explicitly makes a trade-off between confidence about the precise position of the wavefronts for the robustness of higher ensemble spreads, this robustness never materializes in practice; the Autonomous experiment spread is comparable to that from the Wavefront Uncertainty experiment.

\section{Discussion}

In all results, the unobserved interior layers that are not driven by the LETKF analysis ensemble update form the dominant contribution to the reconstruction error, as expected from previous results using surface observations generated from model runs~\cite{marcotte2021robust}.
However, previous experiments focused on dynamics that are autonomous and (in the large scale) structured by topological features, i.e., the dynamics of scroll waves are organized around their filaments; in the present work we have focused on experimental results that are necessarily non-autonomous and driven by external stimuli, as these conditions are endemic to experimental investigations of cardiac tissue excitation. 
In the present scenario, all the dynamics are driven by the observations and thus the LETKF update of the analysis ensemble, save the Free-Run experiment which corresponds to the absence of any observations, without such topological constraints on the dynamics.

\subsection{Model error interactions with the assimilation}

Several considerations have an outsized impact on the accuracy and stabilty of the reconstruction for this dataset.
The first is the model error due to, primarily, misparameterization.
For example, when the refractory period is sufficiently long, then subsequent excitations generated by the LETKF develop into conduction block due to the refractory behavior of the model.
The blocked wavefronts then prevent accurate reconstruction in the next assimilation step.
Additionally, the propagation speed of the wave is important to match in the model parameterization; slow waves rely on the assimilation scheme to excite the tissue ahead of the wavefront to match the observations, while fast waves rely on the assimilation scheme to suppress the propagation of waves in the model to match the observed wavefront.
That the asymmetry between the ignition and quenching problems has significance for the reliability of the state reconstruction is surprising, at first glance.
However, this effect mirrors our understanding of critical transitions for excitable media: the ignition problem relies only on the local threshold response of the excitable medium, while the quenching problem is sensitive to the particular pattern of the excitation and is thus a more nonlinear process.

The parameterization of the model can also affect the dynamical range of the ensemble states, such that the analysis may drive the fields beyond the range of the observations, i.e. `overshoot'.
In our present problem, the action potentials of the Fenton-Karma model are only weakly bounded to the same domain as the observations, i.e. $0 \leq u \leq 1$, whereas the assimilation step utilizes a strict bound for the analysis states $0 \leq u^a \leq 1.5$.
Due to the short model integration times ($T_a=2.0$~ms) and the overshoot of the analysis step, the state cannot relax sufficiently quickly to account for the overshoot, resulting in anomalously large field values, which likewise must be accounted for in the next assimilation step.
This model-assimilation feedback cycle presents further opportunity for assimilation filter investigations.

{One possible way to improve the robustness of the reconstruction process is by extending the reconstruction to the model parameters in addition to the state vector.
This approach is especially useful for model (parameter) identification in the absence of a known model (parameter set).
We have opted to reserve this avenue of investigation for future work, both because a published model parameter set for this dataset and model exists~\cite{barone2020efficient} and because our stochastic model inflation techniques have previously been shown to be adept at accounting for model errors (specifically those arising from mis-parameterizations)~\cite{marcotte2021robust}.
A thorough study of the identification of model parameters for cardiac electrical excitation models using data-assimilation techniques may provide useful advances for clinical applications.}

\subsection{Observation interactions with the model dynamics}

Observation processing is also a more subtle art than it first appears.
Optical mapping recordings effectively blur the position of the wavefront by recording fluorescence not only from the top-most layer of cells in the tissue, but the transferred fluorescing of excited layers of cells below the surface.
This blurring manifests as, instead of a sharp wavefront, a leading edge that smoothly transitions from quiescent to excited across a width as large as $1$~cm. 
{Numerical post-processing to improve the signal-to-noise ratio of the data is typically achieved with a spatiotemporal convolution, which exacerbates this smoothing of the sharp features of the wavefront into a distribution of marginally excited cells. However, as the experimental data we use during constant pacing reaches steady state, stacking was used~\cite{uzelac2015robust}, which increases the signal-to-noise ratio without further spatial smoothing. }
This distribution of marginally excited cells cannot be reconstructed by a threshold excitable model---we rely on the assimilation scheme to effectively interpret the steep wavefronts generated by the ensemble models for a given smooth observed wavefront.
{We have considered the blurring of the state observations $\obs{b}$ by an observation operator which does not merely sample the field at particular positions, but computes the local average of the field through a fixed-width box-filter convolution.
We anticipated that the convolution length scale would compete with the length scale of the localization process and likewise the convolution filter would compete with the filtering step of the LETKF, and expected an overall reduction in informational content about the ensemble as the elements of the $l$-dimensional $\obs{b}$ are now substantially more correlated, harming the reconstruction by introducing an error into the ensemble observations which is essentially uncontrollable.
We found that, in practice, the difference from the sampling based observation operator used in this work is minimal---cf. Figure~S3 
in the Supplement for further details and results.}

Likewise, the estimation of observation uncertainty for cardiac experimental recordings is an important topic that we have treated only superficially in this work.
As the optical-mapping recordings {and post-processing } involve a spatial averaging that essentially smooths the steep wavefronts endemic to real excitable models, we may heuristically assign a nonlinear uncertainty to the observation data based on the approximate value of the observation, as in Section~\ref{ssec:wunc}. 
Because there are effectively three `slow' states observed in the data for the transmembrane potential---quiescent, marginally excited, and fully excited---and we suppose the middle state is an apparent figment due to the measurement scheme, we may estimate the probability that the observation of a cell in that state should remain in that same state by the next observation time, equivalently, assimilation interval, and interpret this proportionality factor for the uncertainty of the observed transmembrane potential.
For the Fenton-Karma model, we may map these observed slow states based on the range of $u$ and model parameters, specifically $u_c < u_i(t) < u_c^{si}$ representing the marginally excited state. 
The uncertainty of observations outside this range should be determined by the noise floor of the observations.
As the \emph{true} values of $u_c$ and $u_c^{si}$ are undefined, but may be assumed to be close to those used in this work, the smooth extension to all observation values is needed; heuristics for nonlinear observation uncertainty depending on the state lead to a lesser weighting (higher uncertainty) for observations of the wavefront and waveback.
Precisely constructing these estimates requires building a statistical model of the observations, which is beyond the scope of this work.
Our initial experimentation based on heuristics that smoothly increase the uncertainty of observations in the marginally excited region suggest that this subtle modification of uncertainty can have significant impacts on the accuracy and robustness of state reconstructions.

{Finally, while we have focused on a particular dataset for the reconstruction task, we have computed reconstruction with related data ($\mathrm{BCL} = 234\pm 2$~ms) and different model parameter sets.
For a slower BCL the conclusion is the same: the propagation of excitation information from the observations to the ensemble state vectors is slow and requires consistent work from the LETKF to correct, which manifests as peaked errors in the surface reconstructions which match each new excitation.
For alternative model parameters, the periodicity is still matched but the difference in wave shape results in significant increases in reconstruction error.
These experiments suggest that our approach is robust to peculiarities of the dataset and model specification.
Further details are available in the Supplement, see Figures~S4-S8.} 

\subsection{Assimilation interactions with the observation uncertainty}

The specification of the LETKF includes several numerical parameters that we found to influence the efficacy of the assimilation program.
The first is a floating-point parameter named `gross error', designated by $\sigma_g$, which controls whether an observation is sufficiently deviant from the background mean $\ensmean{b}$ to be truncated, with the goal of maintaining a `light touch' for the assimilation and enhancing the stability of the scheme in the presence of large-deviations from the observations. 
For an observation-uncertainty pair $(\obs{o}_i, \unc_{i})$ and corresponding value of the background mean field $\overline{u}_i^b(t)$, if $|\obs{o}_i(t) - \overline{u}_i^b(t)| > \sigma_g \, \unc_{i}$, then the corresponding element of $(\obs{o} - \overline{H}(\ens{b}))$ is overwritten by $0$, i.e., the update of the analysis ensemble is invariant with respect to this observation.
As the range of $0 \lesssim u^b \lesssim 1$ and the minimum of the uncertainty of the surface observations $\underline{\unc}  = 0.05$ are fixed, when $\sigma_g \gtrsim 20$ every observation is used in every assimilation step. 
However, we find that setting the parameter below this saturation value (e.g., $\sigma_g = 10$) can significantly affect the results, producing maximal surface reconstruction errors nearly $25\%$ lower than those in the saturated case ($\sigma_g = 20$). 
For this reason, we set $\sigma_g = 10$ for all experiments in this work.
In program logs, it is clear that fewer than $5\%$ of observations are ignored by assimilation step $50$ ($100$~ms) for the autonomous experiment, which suggests that we are only filtering truly extreme values, and only in the initial iterations of the assimilation.
In principle, controlling this parameter is akin to inflation -- it is a measure of our relative confidence in the observations and the background ensemble members.
It is not clear \emph{a priori} what value this parameter should take, and because it affects the accuracy of the reconstruction, characterizing its effects should be a priority in future investigations.

{Further, we have used $\underline{\unc} = 0.05$ throughout this work, which was estimated from the full-width at half-maximum of the distribution of the observation values.
While we found some sensitivity to the precise value of the uncertainty (e.g., experiments with $\underline{\unc} = 0.025$ and $\underline{\unc} = 0.10$ produced reconstructions with higher $\RMSE{b}$ overall), they are qualitatively similar to the presented results.
A rigorous estimation of the observation uncertainty from the statistical properties of the dataset is beyond the scope of the present work, but may reveal that the accuracy of the reconstruction may be improved through better estimation of the observation covariance matrix.}

In this work we have used `strongly coupled' DA (in analogy to the same term used in domain-decomposed models in weather forecasting) for the analysis update to the state variables for all experiments.
We permit observations of the state $u$ to affect the analysis update of the $v$ and $w$ fields.
In contrast, `weakly coupled' DA communicates information between the state variables only through the model evaluation.
Strongly coupled DA is expected to be able to extract more information from the same observations than weakly coupled DA~\cite{sluka2016assimilating}.
Strongly coupled DA permits more significant corrections to the state for our relatively sparse observation pattern, at the risk of over-driving the corrections to $v$ and $w$ in ways that are atypical for cardiac excitations. 
Restricting the analysis update due to observations of the $u$ variable to the $u$ variable only reduces the efficacy of the assimilation by preserving more ensemble state information in the analysis, while permitting larger ensemble spreads due to the uncontrolled dynamics of $v$ and $w$.
In numerical experiments, we found that in some scenarios we could drive the gating variables to their rest state values during the assimilation, when in model simulations of comparable dynamics the gating variables never attain their rest values.
Open questions include whether the relatively weak coupling of the cardiac model is sufficient to ameliorate some of the challenges to robustness strongly coupled DA is expected to expose, or whether a weakly coupled DA approach unacceptably degrades the accuracy of the reconstructions.
A systematic analysis of strongly and weakly coupled DA in the cardiac context would clarify the effect this choice has on robustness and accuracy and should be a focus of future work.

Relatedly, when we found that the LETKF was over-driving the gating variables and leading to physically unlikely values, we bounded the gating variables between their explicitly known limit values, $0 \leq v^a,\,w^a\leq 1$.
This bounding prevents the assimilation from creating unreachable analysis states that have little to do with the true dynamics---a hazard for what is effectively an initial condition for the model.
Additionally, we found that a similar effect could be found in $u^a$, where some ensemble members would be driven to exceedingly large-valued and hyper-localized corrections of the state leading to anomalously large ensemble spreads (i.e., $\SPRD{a}_u \gtrsim 4.0$), which should not be possible for typical values of $u$.
To suppress this growth, we bounded $-0.1 \leq u^a \leq 1.5$, which has the positive effect of bounding the ensemble spread.
Further, while the limits on $u^a$ are sufficiently strong to bound the ensemble spread, they are not strong enough to prevent overshoot such that $u^a > 1 \ge u^o$.
However, it is presently unclear whether this is an innocuous addendum to the assimilation procedure to account for numerical instabilities in the model-LETKF interaction or symptomatic of a limitation of our approach.
Nor are the constraints on the initial conditions to the model sufficiently well-defined mathematically to say these limits need never be adjusted for this {or another} model.

The observation localization is controlled by two parameters for space and one for time; the latter is set so that only the current time affects the assimilation as this is compatible with the clinical application and is computationally efficient.
The spatial localization scales are isotropic, and we have not investigated the choice of different localization scales or anisotropies.
In principle, given the density and low noise of surface observations, it may be computationally advantageous for their influence to extend through the depth of the tissue while retracting their overlap on the surface.
However, this perspective is difficult to justify physically---indeed, it assumes a surface-dominant dynamics of the interior that is, at an abstract level, the phenomenological question we are considering in this work.
Physically, the influence of each observation may extend to the spatial region that may be excited by the propagation of excitation information in the time between observations---that is, $\sigma_{o,i} \propto c_i T_a \propto \sqrt{D_i}$ for a quiescent state---making it wholly anisotropic, and, in fact, shortening the extent of influence into the thickness of the tissue compared to the effect along the surfaces, as $D_z = D_\perp \ll D_\parallel$, by design.
The corresponding fiber-informed localization scales are then related by the inverse conductivity anisotropy ratio, $\sigma_{o,\parallel} = (D_\parallel/D_\perp) \sigma_{o,\perp}$, and vary with spatial position.
Likewise, this scaling informs the observation distribution---while we are limited to surface recordings of electrofluoresence, a consideration of localization from the properties unique to cardiac excitation may benefit the cardiac researcher.

\subsection{Accuracy measurement in the absence of `truth' states}

Finally, we have focused on the surface error of the state reconstructions and appealed to CRPS for a measurement of ensemble consistency in the absence of interior observations or a `truth' state.
The choice of this measure is motivated by its convergence properties with respect to ensemble size and its simple interpretation rather than its particular relevance to cardiac state reconstruction.
Alternative error measures have been used in other works, e.g., threshold-based error~\cite{hoffman2020sensitivity}, which makes reference to an unknown truth state, a prescribed threshold value, and knowledge of the dynamical properties of the model.
Constructing a fair (in the sense of CRPS) measure of unknown state feature error requires the construction of a statistical model of the dynamics of the state itself, against which we might compare the expectations of the reconstructed state. 
This effort would also aid in physically motivated estimation of observation uncertainties, as mentioned above.

{The creation of a statistical model of the dynamics for the state reconstruction problem is precisely the approach taken by researchers in the machine-learning literature, which has shown promising results.
Ref.~\onlinecite{stenger2023reconstructing} investigated the use of a long short-term memory recurrent neural network, autoencoder, and diffusion artificial neural network (ANN) models for the 3D reconstruction problem for a simple excitable model, while varying the history available to the model and the layer depth of the reconstruction.
The central results of this approach are encouraging, especially for shallow depths (close to the observations) and long histories.
Notably, considering the parsimony of their observations ($l/N \approx T\cdot 8.3\times10^{-3}$, for $ 1\leq T \leq 32$) compared to this work ($l/N \approx 1.5\times 10^{-3}$), the authors found similar reconstruction limitations to our approach for a remarkably different dynamical model and geometry.
Ref.~\onlinecite{lebert2023reconstruction} investigated the reconstruction problem for the Aliev-Panfilov model using a U-net architecture ANN with similar considerations to the informational content of the observations as our work.
In particular, the authors found that simultaneous observations of both the top and bottom layers of the excitable medium produced more reliable reconstructions than single-surface observations, but that `projection' observations---those integrating over the depth of the tissue---worked even better.
Their results using the U-net ANN architecture in the `laminar' dynamics regime (an isolated scroll wave with size comparable to the depth of the tissue) are comparable to previous \emph{in silico} results for a similar dynamical pattern produced by a different cardiac excitation model in conjunction with LETKF and stochastic inflation.~\cite{marcotte2021robust}
The authors also found that the encoding of depth information into the observations was essential for the success of their methods, which limits its application to ventricular fibrillation.~\cite{lebert2023reconstruction} }

{The combination of statistical model (machine-learning) and dynamical model (data assimilation) approaches to reconstruction is an active avenue of research.
One of the central pragmatic limitations of ensemble data assimilation approaches is that storage requirements grow as $O(MN)$ while computation (for LETKF~\cite{hunt2007efficient}) grows as $O(M^2N)$; replacing some of the ensemble members with a pre-trained machine learning model could reduce both costs.
Likewise, machine-learning models require training, and do not offer an estimate of their prediction uncertainty; coordination within a data assimilation framework could alleviate the latter issue with existing infrastructure, while hot-swapping dynamical models with machine-learning models could provide relief for the former.
In particular, we have found that it may take up to $250$~ms for the ensemble to settle into a good approximation of the true uncertainty of this system; machine learning may provide an accelerated convergence path through better estimates for the initial state covariance, or ensemble state initial conditions, based on extensive training data.}

\section{Conclusion}

In this work we have performed several assimilation experiments using physical data with variations to the model and observations.
We have found that the assimilation of experimental observations for cardiac reconstruction brings unique challenges compared to the assimilation of model-produced observations.
The leading contributors to difficulty for state reconstruction are the type of dynamics under study, i.e., non-autonomous stimulated dynamics, model error due to mis-parameterization, the presence of synchronizing inputs (i.e., the stimulus current), and the estimation of uncertainties for experimental observations.

When the model generates fundamentally different types of dynamics compared to that described by the experimental data, we rely on significant corrections from the LETKF.
These large corrections, in turn, increase the likelihood of generating physically unrealizable states.
In the cardiac context, unphysical behavior may manifest as an excitation that spreads significantly faster than the model permits for our parameter setting---thus, the reproduction of the wave speed by the model is an important feature for the reproduction of the state dynamics.

An important issue, only briefly touched upon in this manuscript, is the tuning of model parameters to account for data features.
Especially relevant for this dataset is the ability to reproduce alternans -- both in the duration (APD) and amplitude (APA) of excitations.
While APD alternans are readily reproducible with the Fenton-Karma model, APA alternans are not a common feature in this model. 
Modifications to the dynamics or model parameters to reproduce APA alternans in the Fenton-Karma model may be designed, but are outside the scope of the present work.

Model error forms a dominant contribution to the reconstruction error on the surfaces, and has significant effect on the interior spread of the ensemble and the confidence in the reconstruction in the interior.
The $\CRPS{a}$ may give an overly optimistic estimate of the confidence of the reconstruction accuracy in the interior due to synchronization from the periodicity of the observations over time.
Further methods of assessing reconstruction performance without recourse to observations are necessary.

Our data assimilation scheme, while effective for reconstructing the state from model-derived observations with the same distribution pattern in use in the present study, struggles with the experimental data in use in this work. 
We suspect that this difficulty reflects a fundamental difference between our previous three-dimensional reconstruction efforts and the current dataset, which is the non-autonomous nature of the observed dynamics.
Data assimilation experiments with \emph{experimental} data showcasing autonomous dynamics---i.e., a spiral or scroll wave---would present an interesting and informative extension of this study. 

In conclusion, we have found that we can reliably reconstruct the surface dynamics of a driven excitable state from experimental observations on the surfaces, but that reliable reconstruction of the interior requires further efforts to understand our dataset and how it interacts with the dynamics of excitable models.
We have found that while inclusion of explicit modeling of the stimulation current or stochastic effects may yield subtle improvements to the reconstruction, the coupling of the LETKF, observations, and the model is sufficiently close that it does not correct the leading error cause in the reconstruction of this dataset.
On the surfaces, these errors are dominated by poor model fit to the excitation wavefronts and wavebacks.
In the interior, it is the lack of propagation of information from the surfaces, such that the interior becomes a fount of uncertain excitation.
Likewise, we have determined that incorporating additional modeling insight into the features of the observations, whether through the identification of a coherent wave pattern and using it to constrain the uncertainty of the interior dynamics or through modeling of the uncertainty associated with the observations, is subtle and produces new challenges which, at this initial stage, do not appear to improve on the low-information approach.

\section{Supplementary Material and Data Availability \label{ssec:supp}}

We have included two video files in the {online } supplement.
The first, \texttt{auto\_ensmeans.mp4}, displays the dynamics of $\overline{u}^{b}(t,x,y,z)$ and $\overline{u}^{a}(t,x,y,z)$, for $z/d = 0, 0.2, 0.4, 0.6, 0.8, 1.0$, and emphasizes the difference due to the assimilation, $\overline{u}^a - \overline{u}^b$, and the observations $u^o$, for the autonomous model experiment.
The second, \texttt{auto\_b.mp4}, displays the dynamics of  $\overline{u}^{b}(t,x,y,z)$,  $\overline{v}^{b}(t,x,y,z)$, and $\overline{w}^{b}(t,x,y,z)$ over time, for $z/d = 0, 0.2, 0.4, 0.6, 0.8, 1.0$ for the autonomous model experiment.

{We have also included a short supplement which includes results for a number of experiments referenced throughout the text.}

The data that support the findings of this study are available from the corresponding author upon reasonable request. 

\begin{acknowledgments}
We thank Alessio Gizzi for providing access to the data used for this study. 
M.J.H. and E.M.C. were supported by the NSF under Grant Nos. CMMI-2011280 and CMMI-1762803. F.H.F. was supported by the NSF under Grant No. CMMI-1762553. C.D.M., F.H.F., and E.M.C. were supported by the NIH under Grant No. 1R01HL143450-01.
This research was supported in part through research infrastructure resources and services provided by the Partnership for an Advanced Computing Environment (PACE) at the Georgia Institute of Technology, Atlanta, Georgia, USA~\cite{PACE}. 
This work has likewise made use of the Hamilton HPC Service of Durham University.
All figures in this report were generated using the \texttt{Makie.jl}~\cite{DanischKrumbiegel2021} software package.
\end{acknowledgments}

\bibliography{DinDA}

\end{document}


\section{Supplemental Materials}

\subsection{Reconstructing Action Potentials}
We include here temporal traces of the reconstruction and coordinating observation trace for several numerical experiments presented in the main text.
The location of the trace is selected randomly from the coordinates of the observations for each experiment. 

\begin{figure}[htpb]
    \centering
    \includegraphics{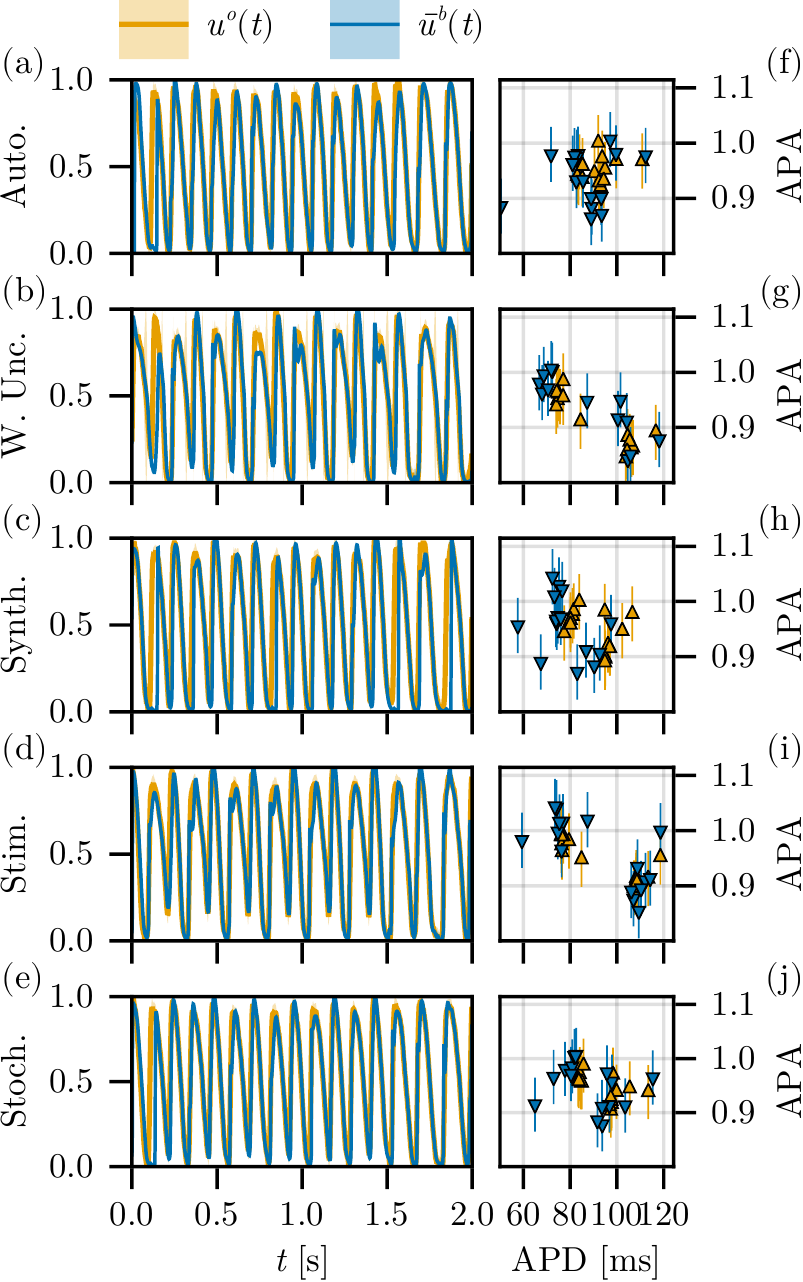}
    \caption{Pairs of (a-e) temporal traces of the reconstruction $\overline{u}^b(t)$ and observation $u^o(t)$ at randomly selected locations for experiments presented in the main text, and (f-j) the distribution of APD and APA for both traces ($u_\mathrm{thr} = 0.2$). }
    \label{fig:traces}
\end{figure}

In all experiments, the reconstruction reliably recovers the action potential (AP) shape after a short transient of less than two cycle lengths ($< 250$~ms), cf. figure~\ref{fig:traces}(a-e).
The dominant error of the AP shape occurs near the apex, where over- and undershoot of the reconstruction lead to poor estimates for the APA.
In contrast, since the estimates for the APD use a threshold which is smaller than the apex, the estimates are consistent between the observation and reconstruction.
The APD and APA calculations used to produce figure~\ref{fig:traces}(f-j) use a threshold  $u_\mathrm{thr} = 0.2$ to avoid reporting APD + BCL for the long waves which do not reach quiescence before re-ignition.

\subsection{Estimation of the minimal observation uncertainty}

A rigorous calculation of the minimal observation uncertainty is beyond the scope of this work.
However, we estimated the minimal observation uncertainty $\underline{\eta}$ from the full-width at half-maximum of the distribution of delayed differences in the dataset, using the assimilation interval for the delay time, $\tau=2.0$~ms.

\begin{figure}[htpb]
    \centering
    \includegraphics{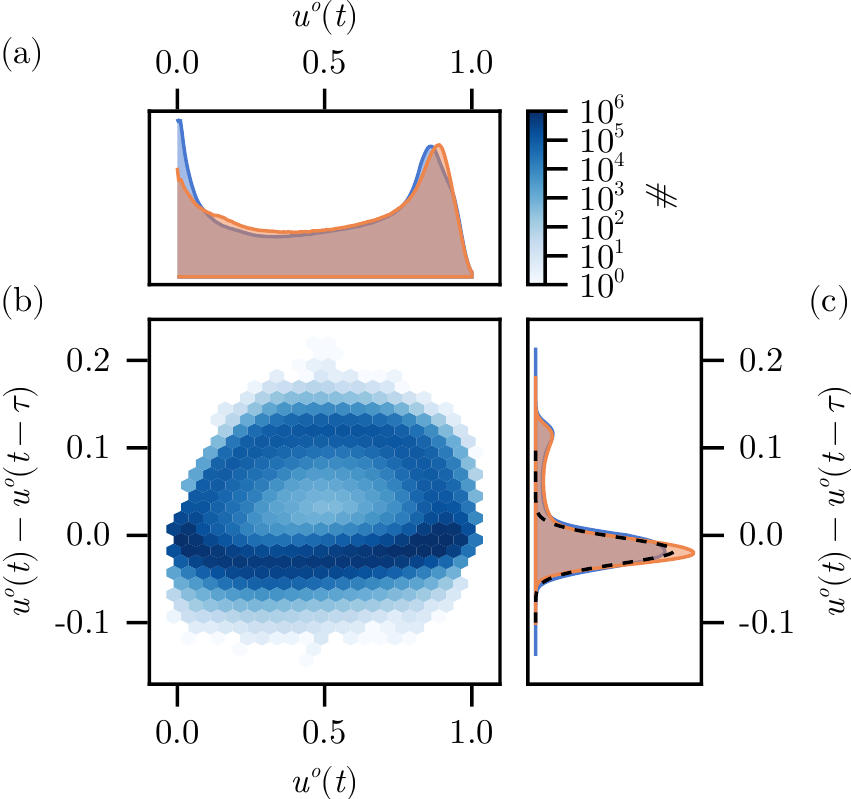}
    \caption{(a) Distribution of field values in the dataset, (b) distribution of values in the dataset against the assimilation interval time-delayed difference, and (c) the distribution of time-delayed differences in the dataset corresponding to the assimilation interval. The baseline uncertainty of the observations is estimated from the full-width at half-maximum of the Gaussian peak occurring near zero (black dashed curve). }
    \label{fig:obs-stats}
\end{figure}

In figure~\ref{fig:obs-stats}(b) we show the distribution of the observations against their assimilation-interval-delayed differences with a resolution of $(0.05, 0.05/3)$ on the $(u^o(t), u^o(t)-u^o(t-\tau))$ axes, respectively.
Beginning from the rest state $(0,0)$, the dynamics progress around the apparent loop in the clockwise direction.
The marginal distribution of $u^o(t)$ and $u^o(t)-u^o(t-\tau)$ is shown in figures~\ref{fig:obs-stats}(a,c), respectively.
We estimate the full-width at half-maximum of the distribution in (c) using a Gaussian
with standard deviation $\sigma = \underline{\unc}/(2\sqrt{2\ln2})$ for the peak near $0$, with $ \underline{\unc} \approx 0.05$, shown as a dashed black curve for $u^o(t) - u^o(t-\tau) \in [-0.1,0.1]$.
This value likewise makes the identification of quiescent observations between $0 \leq u^o \leq 2\underline{\unc}$ more reliable.

\subsection{Observation Operator Local Averaging}

In the body of this work, the observation operator $H(\ens{b}) = \obs{b}$ is the action of a large, sparse, rectangular matrix, to reduce the $N$-dimensional state vector to the $l$-dimensional observation of the state, $l \ll N$.
The operator strictly samples the $u$-field at particular positions.
Meanwhile, the observation data from the experiment is filtered twice: first by the transmission of subsurface fluorescence, and second by post-processing with a spatiotemporal kernel for an improvement of the signal-to-noise ratio of the recording.
The effect is to smooth the dataset by reducing the influence of outlier signals, which also removes the resolution of truly sharp features in the system from the data.

\begin{figure}[htpb]
    \centering
    \includegraphics{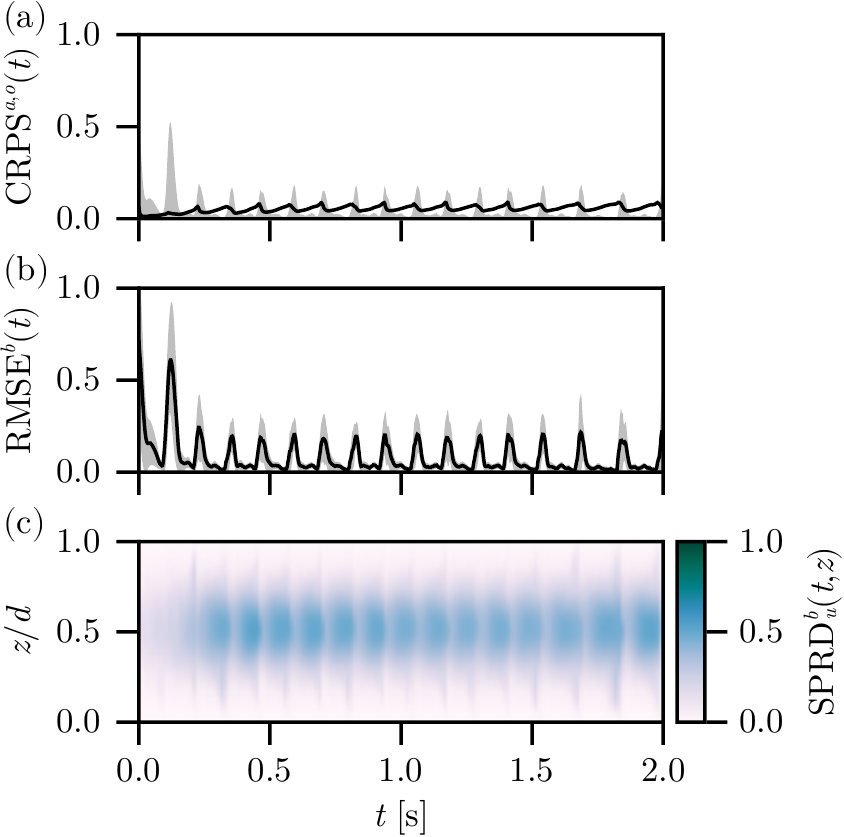}
    \caption{Autonomous model reconstruction results with the spatial convolution observation operator reporting the (a) $\CRPS{a,o}(t)$, (b) $\RMSE{b}(t)$, and (c) spread through the depth $\SPRD{b}_u(t,z)$. }
    \label{fig:supp_obs_avg}
\end{figure}

To this end we have extended the observation operator; rather than $H(\ens{b})$ returning the value of the field at a particular index (position) $u_i^b$, we instead compute and return the local average in a $p\times p$ patch centered on the index (position), where we have chosen $p=5$ so that observations (in our default surface distribution) sample the entire surface in aggregate, but each observation does not include any other observation position in its average.
This is equivalent to a spatial convolution with a two-dimensional box filter.
Results of this filtering observation operator are shown in figure~\ref{fig:supp_obs_avg}, corresponding to the autonomous reconstruction problem (cf. Section~III.A.2 in the main text).
The results are marginally different from the canonical sampling observation operator, with minor increases in the $\RMSE{b}(t)$ and marginal decreases in the ensemble spread, overall.

\subsection{Stimulus Period Robustness}

\begin{figure}[htpb]
    \centering
    \includegraphics{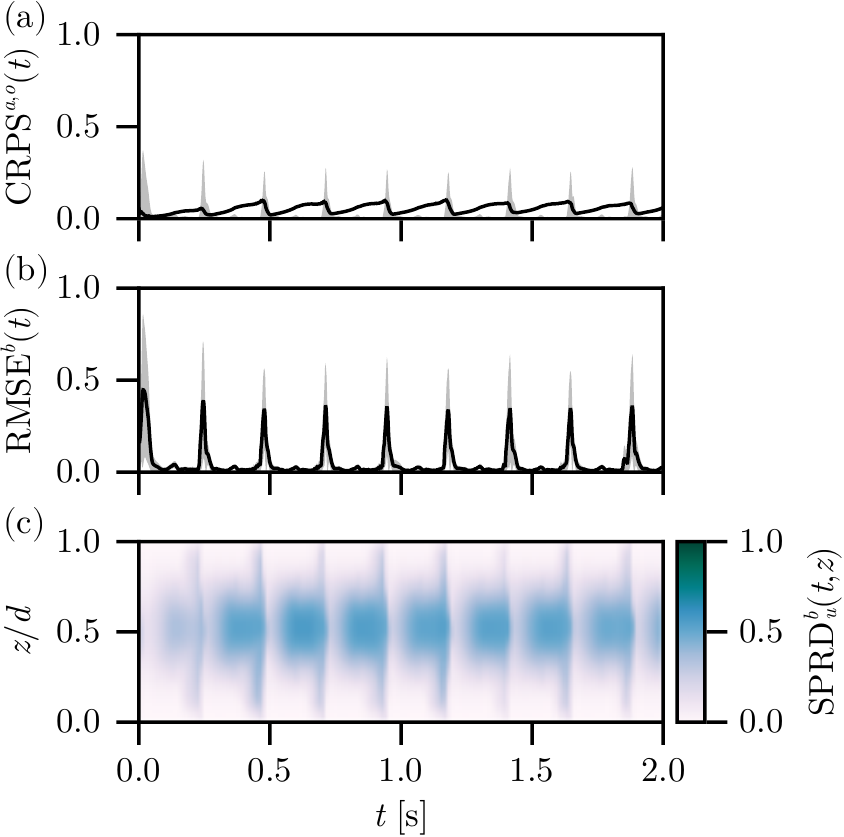}
    \caption{Reconstruction results with the dataset with $\mathrm{BCL} = 234\pm2$~ms, reporting the (a) $\CRPS{a,o}(t)$, (b) $\RMSE{b}(t)$, and (c) spread through the depth $\SPRD{b}_u(t,z)$. }
    \label{fig:auto_92}
\end{figure}

In addition to the robustness of our approach with respect to model parameters, we applied the reconstruction to additional datasets with longer cycle lengths, which produce substantially slower dynamics and less pronounced alternans.
However, despite the relatively calmer dynamics, the assimilation becomes substantially more sensitive to incorrect AP shapes.
We consider an autonomous model reconstruction result for a significantly longer $\mathrm{BCL} = 234\pm2$~ms in figure~\ref{fig:auto_92}. Both the $\CRPS{a}(t)$ and maximal $\RMSE{b}(t)$ are significantly increased for this reconstruction.

\begin{figure}[htpb]
    \centering
    \includegraphics{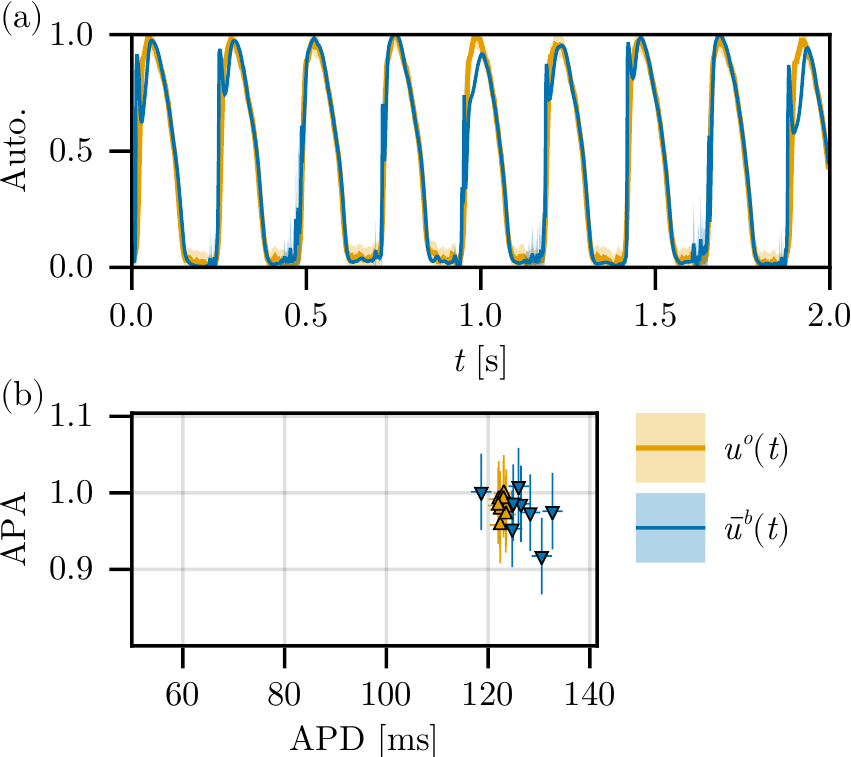}
    \caption{(a) Temporal traces for the observations with BCL=$234\pm 2$~ms and the reconstruction using the autonomous model, and (b) the distribution of APD and APA for the APs in this sequence ($u_\mathrm{thr}=0.2$). }
    \label{fig:auto_92_traces}
\end{figure}

While the deviation in the $\RMSE{b}(t)$ is high-amplitude, it is very short-lived (cf. figure~\ref{fig:auto_92}(b)), and the AP shape is reliably recovered for this dataset (cf. figure~\ref{fig:auto_92_traces}(a)).
As the model is able to fully recover between excitations, an entire mechanism of nonlinear feedback between the model, observations, and the LETKF is no longer in effect.
As expected for simpler dynamics with longer BCLs, our approach performs remarkably well both in detail and in aggregate.
Notably, often the reconstructed AP shape in figure~\ref{fig:auto_92_traces}(a) exhibits a pronounced spike-and-dome structure, a structure which is not typically exhibited for this model but appears in numerous ionic models of cardiac excitation.
The appearance of this structure is why depictions of the AP for the reconstruction must be approached with caution, as this is likely due to an over-eager excitation by the model followed by LETKF correction of the ensemble states, rather than a statistical feature of the data being reproduced by the reconstruction. 

The APD and APA distribution is well-matched for this trace, cf. figure~\ref{fig:auto_92_traces}(b); the former is expected based on the shorter BCL results, and the latter appears to be due to the longer period between excitations, lessening the impact of APA alternans and thus better tuning to the attainable AP morphologies for the Fenton-Karma model.
That is, while the dynamics is well within the 1:1 regime, the reconstruction is better able to correct for the AP shape because the entire ensemble is guaranteed to have the same periodicity and phase, whereas in regimes with strong alternans some ensemble members may be on a different beat than others, constraining the available degrees of freedom for correcting the AP shape.

\subsection{Model Parameter Sensitivity}

\subsubsection{Conductivities}

In Ref.~\onlinecite{barone2020experimental} the model parameters are determined (``manually tuned'') and the conductivity of the tissue is estimated using a customized data assimilation approach.
The model parameters are used throughout this manuscript in the reconstruction problem, save for the estimated conductivities, which are unphysically high to account for the very fast propagation of excitation across the tissue.
The values used in this manuscript are $D_{\parallel} = \sigma_{\parallel}/9$ and $D_{\perp} = \sigma_{\perp}/15$, where $\sigma_{\parallel,\perp}$ are the reported estimates for the fiber-aligned and fiber-orthogonal conductivities, respectively.

In the interest of robustness, we tested the unphysically high conductivities (with a reduced model integration time-step $\delta t = 2\times 10^{-4}$) in our reconstruction problem.
As stated in the text, we found that the reconstruction was substantially worse than the Free-Run experiment as without a periodic stimulus current, the ensemble returns to quiescence and is unable to re-excite through the LETKF driving alone.

\begin{figure}[htpb]
    \centering\includegraphics{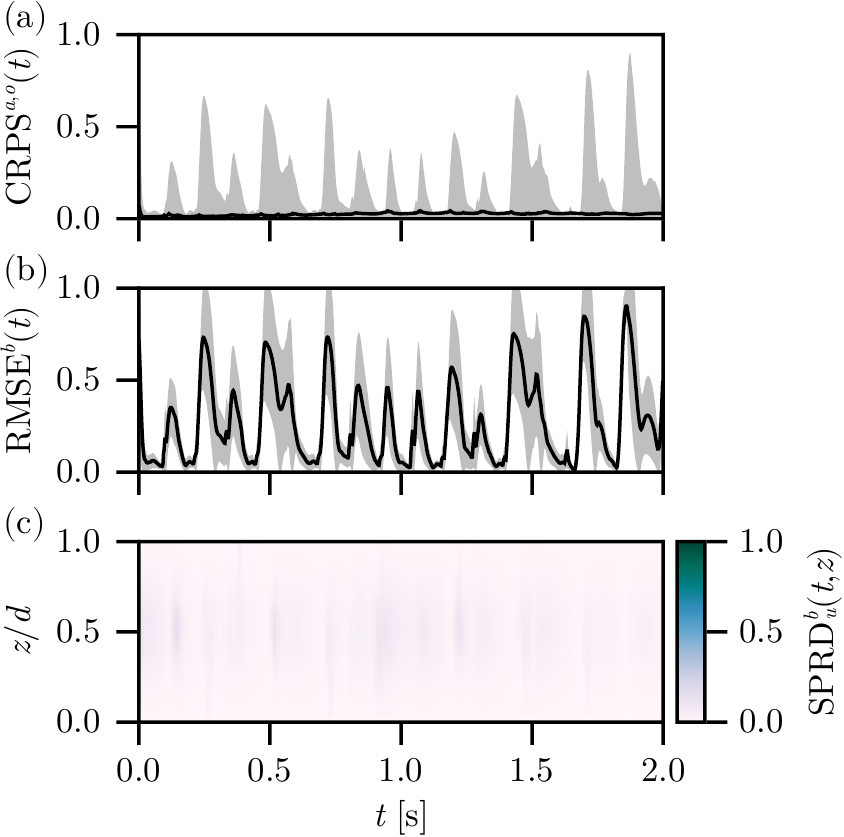}
    \caption{Reconstruction for the stimulated Barone model with $D_{\parallel,\perp} = \sigma_{\parallel,\perp} = (0.009, 0.003)$~cm$^2$/ms from Ref.~\onlinecite{barone2020experimental}, with the (a) $\CRPS{a,o}(t)$, (b) $\RMSE{b}(t)$, and (c) spread through the depth $\SPRD{b}_u(t,z)$. }
    \label{fig:fast_barone_stim}
\end{figure}

In conjunction with the stimulus current, we were able to match the baseline frequency of the excitations in the observations and consistently ignite the ensemble states.
Figure~\ref{fig:fast_barone_stim} depicts the reconstruction problem for the stimulated `fast' model with unphysically large conductivities.
Clearly this reconstruction is worse in every measure---indeed, even with these high conductivities, the ensemble can not match the wave speed in the observations, due to consistent interaction with refractory tissue.

\subsubsection{Excitable model parameters}

Finally, we turn to the excitable model parameters themselves.
As the Fenton-Karma model is sufficiently flexible to fit large space of possible restitution curves, we used three different parameter sets to compare the reconstruction sensitivity to the excitable model parameters, summarized in Table~\ref{tab:model_params}.
The Beeler-Reuter (BR) and Guinea Pig (GP) parameter sets are adapted from Ref.~\onlinecite{fenton1998vortex}.

\begin{table}[htpb]
    \centering
    \begin{tabular}{c|S|S|S|c} 
        Parameter & BR & GP & {Barone} & units \\
        \hline\hline
        $\tau_{si}$      &   30.0    & 22.0  & 127   & [ms] \\
        $\tau_{v1}^{-}$  &   1250    & 333   & 38.2  & [ms] \\
        $\tau_{v2}^{-}$  &   19.6    & 40.0  & 38.2  & [ms] \\
        $\tau_{v}^{+}$   &   3.33    & 10.0  & 1.62  & [ms] \\
        $\tau_{w}^{-}$   &   41.0    & 65.0  & 80    & [ms] \\
        $\tau_{w}^{+}$   &   870     & 1000  & 1020  & [ms] \\
        $\tau_{d}$       &   0.25    & 0.12  & 0.1724& [ms] \\
        $\tau_{o}$       &   12.5    & 12.5  & 12.5  & [ms] \\
        $\tau_{r}$       &   33.0    & 25.0  & 130   & [ms] \\
        $k$              &   10.0    & 10.0  & 10    & \\
        $u_c$            &   0.13    & 0.13  & 0.13  & \\
        $u_v$            &   0.04    & 0.025 & $*$   & \\
        $u_c^{si}$       &   0.85    & 0.85  & 0.85  & \\
    \end{tabular}
    \caption{Fenton-Karma model parameter values used in dynamical model for the reconstruction of experimental data. $^*$Note: as $\tau_{v1}^- \equiv \tau_{v2}^-$, the switching parameter $u_v$ is unspecified in Ref.~\onlinecite{barone2020experimental}.}
    \label{tab:model_params}
\end{table}

\begin{figure}[!h]
    \centering
    \includegraphics{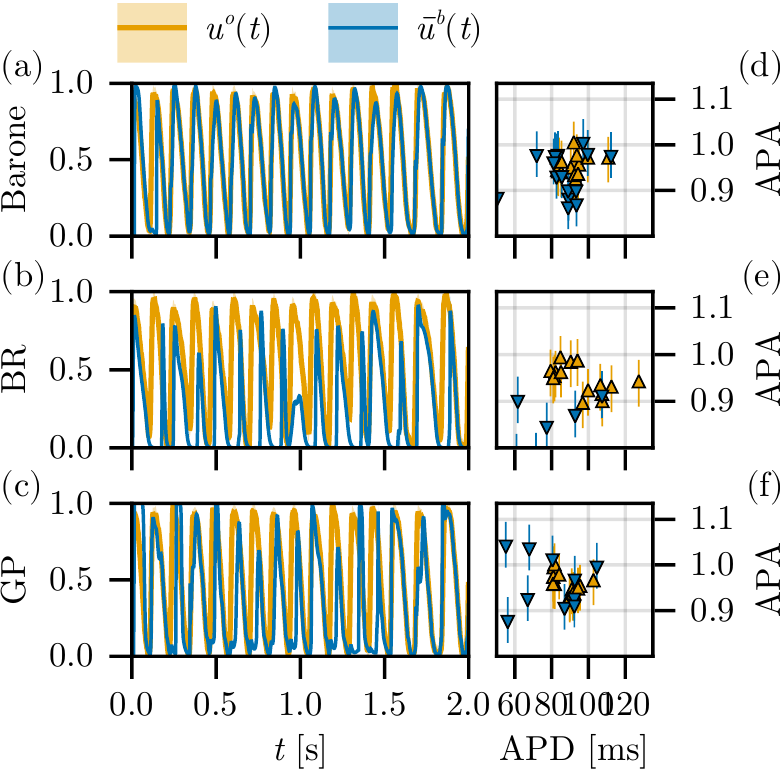}
    \caption{Pairs of (a-c) temporal traces of the reconstruction $\overline{u}^b(t)$ and observation $u^o(t)$ at randomly selected locations for the Barone \emph{et al.}, BR, and GP parameter sets, and (d-f) the corresponding distributions of APD and APA ($u_\mathrm{thr} = 0.2$). }
    \label{fig:param_traces}
\end{figure}

\begin{figure*}[htpb]
    \centering
    \includegraphics{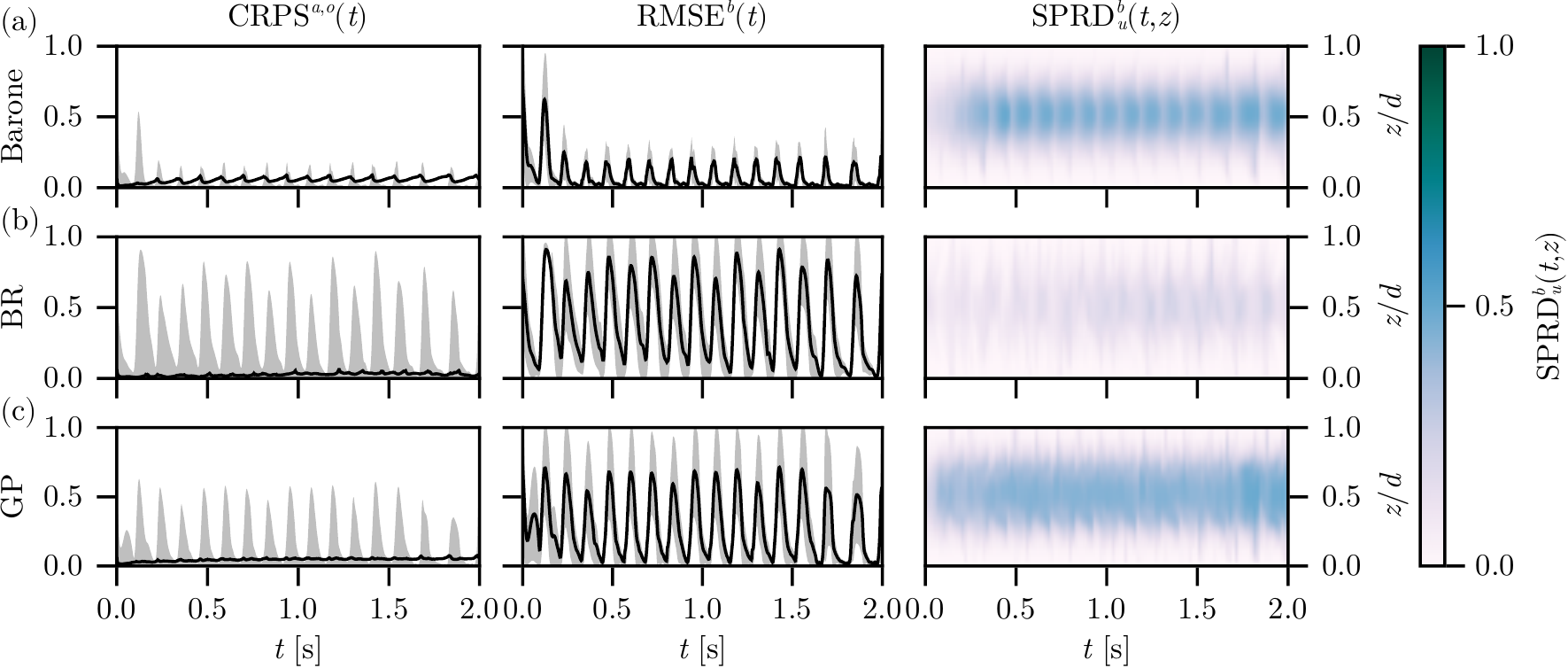}
    \caption{Reconstruction results for the autonomous model using the (a) Barone et. al, (b) Beeler-Reuter, and (c) Guinea Pig parameter sets. }
    \label{fig:auto_param}
\end{figure*}

In figure~\ref{fig:param_traces}(a) we reproduce the temporal traces of the observations and reconstruction using the Barone \emph{et al.} parameter set for a randomly selected position covered by the observations, and show the APD and APA distributions for both signals in figure~\ref{fig:param_traces}(d).
The BR and GP model parameters predominantly impact the AP shape, driving the larger $\RMSE{b}(t)$ in figure~\ref{fig:auto_param}(b,c).
The AP shape is not reliably recovered for the reconstructions with the BR and GP parameter sets, cf. figure~\ref{fig:param_traces}(b,c), respectively.
As a result, both the BR and GP models perform worse at approximating the APD and APA distributions of the observations, cf. figure~\ref{fig:auto_92_traces}(e,f), respectively.
This indicates that the Barone \emph{et al.} parameter set adequately models the underlying true dynamics, despite localized surface reconstruction errors during each new excitation.

We computed autonomous data assimilation reconstructions for each new parameter set to understand the impact of the excitable model parameters on the reconstruction, and summarize these results in figure~\ref{fig:auto_param} in which we (a) reproduced the the baseline autonomous reconstruction outputs with the Barone \emph{et al.} parameter set used throughout the main text, and consider the reconstruction using (b) Beeler-Reuter and (c) Guinea Pig parameter sets, reporting $\CRPS{a,o}(t)$, $\RMSE{b}(t)$, and the spread through the depth $\SPRD{b}_u(t,z)$.
Not only is the $\RMSE{b}(t)$ substantially larger (worse) for the (b) Beeler-Reuter (BR) and (c) Guinea-Pig (GP) parameter sets, the change in excitability parameter $\tau_d$ yields dramatically different AP shapes and conduction velocities; this likewise changes the steady-state distribution of ensemble spread through the depth.
For example, slower waves for the BR parameter set propagate into the interior less over the assimilation interval leading to higher erroneous certainties, while the relatively fast GP waves propagate information from the surfaces into the interior.

Of particular note is the change in the $\CRPS{o}(t)$ for the BR and GP parameter sets, cf. figure~\ref{fig:auto_param}(b,c).
In both cases, the background ensemble shows significant disagreement with the observation sequence compared to the Barone \emph{et al.} parameter set, and likewise reaffirms our earlier conclusion from the AP traces that the parameter set used in this work is an adequate model of the true dynamics for this dataset and model. 

\subsection{Deviation from the Autonomous Model Reconstruction}

\begin{figure}[htpb]
    \centering
    \includegraphics{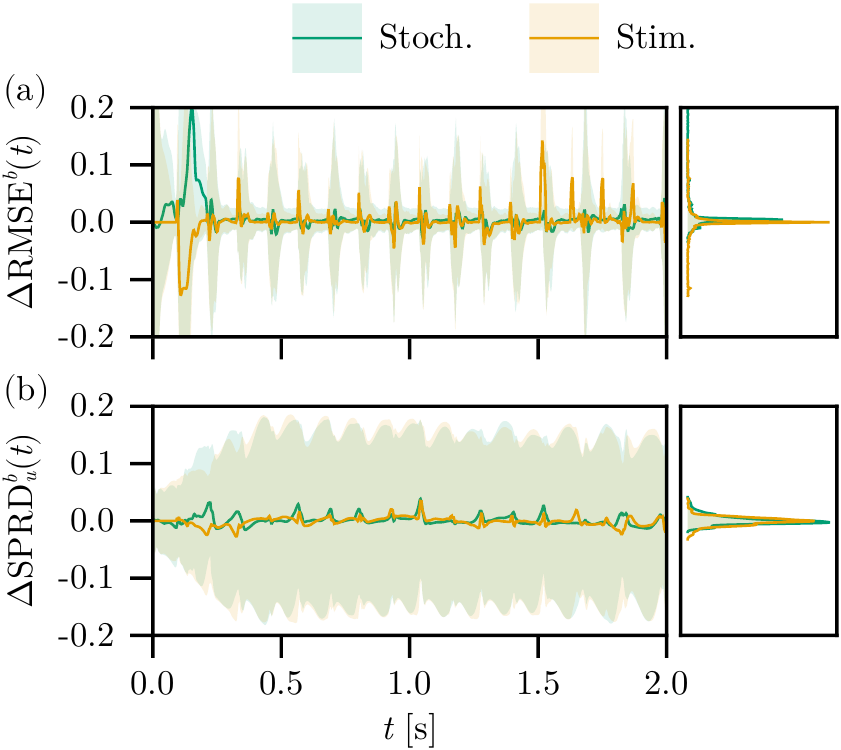}
    \caption{Difference from the Autonomous reconstruction error for the Stochastic and Stimulus model reconstructions; depicting the (a) $\Delta\RMSE{b}(t)$ and (b) $\Delta\SPRD{b}_u(t)$, as well as their marginal distributions.}
    \label{fig:autostimstoch}
\end{figure}

In figure~\ref{fig:autostimstoch}, we compare the \emph{deviation} in the reconstruction error and the reconstruction spread for the Stochastic and Stimulus models, with the autonomous results as a baseline.
Both the Stochastic and Stimulus model reconstructions show minor deviations from the Autonomous results, corresponding to oscillating improvements (for $\Delta\RMSE{b}(t) > 0$) or detriments (for $\Delta\RMSE{b}(t) < 0$) over time.
The reconstruction errors correspond to $\Delta\RMSE{b}(t) = \RMSE{b}_\textrm{*}(t) - \RMSE{b}_\textrm{Auto.}(t)$, and similarly for the spread $\Delta\SPRD{b}_u(t) = \SPRD{b}_\textrm{u,*}(t) - \SPRD{b}_\textrm{Auto.,u}(t)$, where $* \in \{\textrm{Stim.}, \textrm{Stoch.} \}$.
Additionally, we can compare the the error between the background ensemble mean and the observations, denoting this quantity as a vector function $e_{*}(t) = \overline{u}^{b}-u^o \in \mathbb{R}^{l}$, where $l$ is the observation dimension.
The expectation value of the change in the reconstruction error for the Stimulus and Stochastic models over time and space for the error is smaller than the variation in the reconstruction error,
\[
    \textrm{mean}(|e_{*}(t) - e_\textrm{Auto.}(t)|) \ll \textrm{std}(|e_*(t)|),
\]
where $\textrm{mean}$ and $\textrm{std}$ are the standard mean and standard-deviation functions, evaluated over all $1000$ assimilation intervals and all $l$ indices in $e_{*}(t)$, enumerating the observations.
At these scales, figure~\ref{fig:autostimstoch}(b) clearly shows the increased spread of the ensemble for the Stochastic model reconstruction; where peaks in $\Delta\SPRD{b}_{\textrm{Stoch.}, u}(t)$ achieve similar magnitudes as the Stimulus model, but each peak is appreciably non-zero over longer intervals.

\bibliography{DinDA}